\documentclass[
reprint,
amsmath,
amssymb,
floatfix,
superscriptaddress,
twocolumn,
aip,
jcp
]{revtex4-2}
\usepackage{graphicx}
\usepackage{dcolumn}
\usepackage{bm}
\usepackage{color}
\usepackage{amssymb}
\usepackage{amsmath}
\usepackage{xfrac}
\usepackage{mathrsfs}
\usepackage{comment}
\usepackage{mathtools,cancel}

\usepackage[normalem]{ulem}

\usepackage{hyperref}
\usepackage[dvipsnames]{xcolor}
\hypersetup{
    colorlinks=true,
    linkcolor=Blue,
    citecolor=Blue,   
   urlcolor=Blue
   }

\def\la{{\langle}}
\def\ra{{\rangle}}
\def\br{{\textbf{r}}}
\def\tbr{{\tilde{\textbf{r}}}}
\def\Pe{{\rm{Pe}}}
\def\bu{{\hat{\textbf{u}}}}
\def\bR{{\bm{\mathcal{R}}}}

\def\ttt{{\tilde{t}}}

\begin{document}

\title{Chirality, confinement and dimensionality govern re-entrant transitions in active matter}

\author{Anweshika Pattanayak}
\email[Equal contribution:~]{anweshika.pattanayak@tifr.res.in}
\affiliation{Department of Physical Sciences, Indian Institute of Science Education and Research Mohali, Sector 81, Knowledge City, S. A. S. Nagar, Manauli PO 140306, India}
\affiliation{Department of Theoretical Physics, Tata Institute of Fundamental Research, Homi Bhabha Road, Mumbai 400005, India}

\author{Amir Shee}
\email[Equal contribution:~]{amir.shee@uvm.edu}
\affiliation{Department of Physics, University of Vermont, Burlington, VT 05405, United States of America}

\author{Debasish Chaudhuri}
\email[Corresponding author:~]{debc@iopb.res.in}
\affiliation{Institute of Physics, Sachivalaya Marg, Bhubaneswar 751005, India.}
\affiliation{Homi Bhabha National Institute, Anushakti Nagar, Mumbai 400094, India}

\author{Abhishek Chaudhuri}
\email[Corresponding author:~]{abhishek@iisermohali.ac.in}
\affiliation{Department of Physical Sciences, Indian Institute of Science Education and Research Mohali, Sector 81, Knowledge City, S. A. S. Nagar, Manauli PO 140306, India}

\date{\today}

\begin{abstract} 
The non-equilibrium dynamics of individual chiral active particles underpin the complex behavior of chiral active matter. Here we present an exact analytical framework, supported by simulations, to characterize the steady states of two-dimensional chiral active Brownian particles and three-dimensional torque-driven counterparts in a harmonic trap. Using a Laplace-transform approach to the Fokker–Planck equation, we derive closed-form expressions for displacement moments and excess kurtosis, providing a precise probe of non-Gaussian statistics. Our analysis reveals three distinct regimes characterized by bimodal distributions with off-center peaks, Gaussian-like distributions, and weakly heavy-tailed distributions unique to two dimensions. We show that dimensionality plays a decisive role: in two dimensions, increasing chirality suppresses activity and restores Gaussian-like behavior, whereas in three dimensions, torque sustains anisotropic steady states and preserves non-Gaussianity even at high chirality. These behaviors are captured by simple active length-scale arguments that map the boundaries between Gaussian-like and non-Gaussian states. Our results offer concrete experimental signatures — including kurtosis crossovers, off-center peaks, and torque-induced anisotropy — that establish confinement as a powerful tool to probe and control chiral and torque-driven active matter.
\end{abstract}

\maketitle

\section{Introduction}
Active matter remains out of equilibrium by converting energy into self-propulsion, producing persistent motion and breaking time-reversal symmetry~\cite{Schweitzer2003, Mirkovic2010, Ramaswamy2010, Romanczuk2012, Vicsek2012, Marchetti2013, Baconnier2024}.
Natural active systems include motor proteins, bacteria, and cells, as well as larger collectives such as fish schools and bird flocks~\cite{Ndlec1997, Dombrowski2004, Nagy2010, Schaller2010, Katz2011}.
Artificial realizations range from self-propelled colloids and synthetic microswimmers to vibrated polar disks and robotic swarms~\cite{Golestanian2007, Deseigne2010, Palacci2013, Soto2014, Rubenstein2014}.
With intrinsic chirality or an applied torque, active matter forms a distinct class of non-equilibrium systems where self-propulsion, coupled with angular velocity, gives rise to a wide range of complex behaviors~\cite{Sevilla2016,  Caprini2019, Lei2019, Liebchen2022, Bickmann2022, Caprini2023, Debets2023, Ai2023, Caprini2024, Shee2024}.

\medskip

When chirality is present, an active particle traces circular or helical paths, set by an intrinsic angular velocity that often comes from asymmetries in its shape or propulsion mechanism~\cite{Kummel2013, Bechinger2016, Mano2017, Zhang2022, Chan2024, Kaur2025}.
In odd active matter, chirality can arise spontaneously, giving rise to striking properties such as odd viscosity, odd elasticity, and odd diffusivity~\cite{Soni2019, Scheibner2020, Hargus2021, Fruchart2023}.
Chiral motion has been observed in several experimental contexts, such as in active biomolecules~\cite{Jennings1901}, motor proteins~\cite{Loose2014}, microtubules~\cite{Sumino2012}, cells and bacteria~\cite{Brokaw1982, DiLuzio2005, Riedel2005, Lauga2006, Lemelle2010, Nosrati2015, Otte2021, PerezIpina2019}.
In particular, bacteria swimming near a substrate, propelled by flagellar rotation, display clear chiral trajectories~\cite{Otte2021, PerezIpina2019}.
On the theoretical side, chiral motion has been analyzed in both two- and three-dimensional systems~\cite{van-Teeffelen2008, Van-Teeffelen2009, Wittkowski2012, Kummel2013, Volpe2014, Sevilla2016, Lowen2016, Kurzthaler2017, Chepizhko2019, Otte2021, PerezIpina2019, Pattanayak2024, Shee2025Resetting, Santra2025Universal}.

\medskip

The study of chiral motion in active particles is particularly intriguing when these particles are under a confining potential, which biases their position toward the potential minimum~\cite{Caprini2019, Caprini2023}.
Several studies have explored how the medium~\cite{Sprenger2022} and confinement~\cite{van-Teeffelen2008, Ao2015, Caprini2019, Fazli2021, Murali2022} affect the behavior of single chiral active Brownian particles (cABPs).
The combination of chiral dynamics and external confinement creates a rich framework for studying nontrivial steady states and dynamics~\cite{Jahanshahi2017, Caprini2019}.
For example, the interplay between chirality-induced rotational motion and the trapping force can result in various steady state configurations, such as off-center orbiting states and chiral limit cycles~\cite{Jahanshahi2017, Caprini2019}.

\medskip

The interplay between chirality and trap strength modulates effective diffusivity and transport, offering insights into how biological and synthetic systems may exploit similar mechanisms for active transport.
It is therefore important to determine the exact steady states of trapped active particles under torque, identifying parameter regimes that guide experimental realization of chiral active systems.

\medskip

Using an exact analytical framework and numerical simulations, we present a comprehensive investigation of the non-equilibrium steady-state (NESS) properties of two-dimensional chiral active Brownian particles (ABPs) and three-dimensional torque-driven ABPs confined within a harmonic potential.
We employ a Laplace transform-based method, originally developed for semiflexible polymers~\cite{Hermans1952, Daniels1952} and more recently extended to active particle systems~\cite{Shee2020, Chaudhuri2021, Shee2022, Patel2023, Patel2024}, to use the Fokker-Planck equation to  derive exact analytical expressions for arbitrary moments of key observables, including the second and fourth moments of displacement and the excess kurtosis. In particular, the excess kurtosis emerges as a sensitive probe of non-equilibrium behavior in the steady state, capturing how activity, chirality or torque, and trap strength together drive deviations from equilibrium.
Building on prior observations in non-chiral active Brownian particles (ABPs), we show that the inclusion of chiral motion significantly modifies the nature of the re-entrant transition between Gaussian and non-Gaussian states under increasing trap strength.~\cite{Malakar2020, Chaudhuri2021}.

\begin{figure}
\centering
\includegraphics[width=8.75cm]{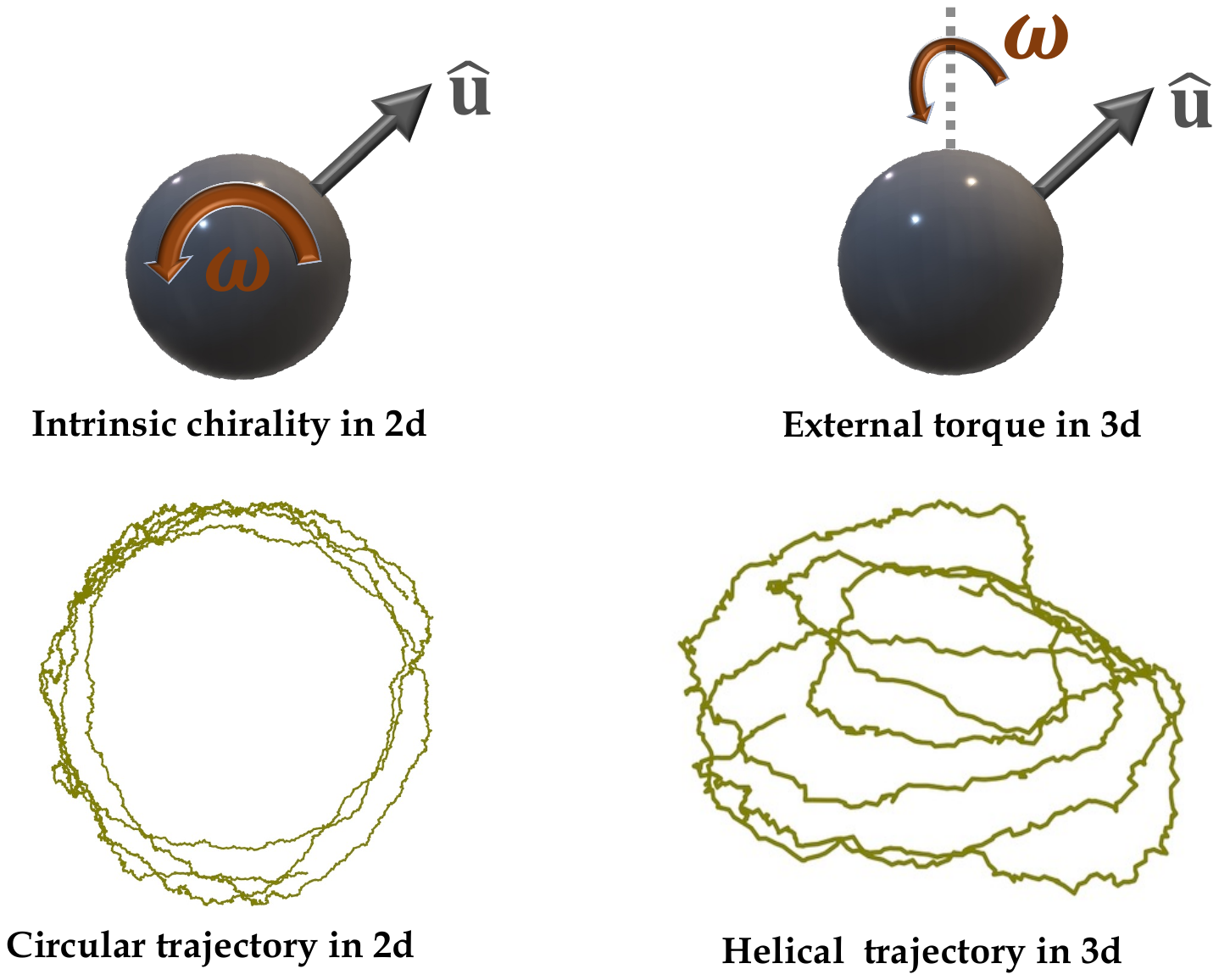}
\caption{
Schematic representation of torque effects in two- and three-dimensional active particles, illustrating their typical steady-state trajectories under harmonic confinement.
}
\label{fig1}
\end{figure}

\section{A trapped Chiral ABP in 2D}
The dynamics of a chiral ABP in two dimensions are described by its position $\br=(x,y)$ and the heading direction  $\bu = (\cos\phi , \sin\phi)$.
In an external harmonic potential $U = k \br^2/2$, the overdamped Langevin dynamics of the chiral particle combine an active velocity $v_0\bu$ with a constant angular speed $\omega$ ( Fig.~\ref{fig1}:
\begin{align}
d{\br} &= v_0\bu dt +\sqrt{2 D}\,d\bm{B}-\mu \br dt\,,\label{eom1:2d}
\\
d{\phi} &= \omega dt+ \sqrt{2 D_r} \, d W\,.
\label{eom2:2d}
\end{align}
Here $\mu = k/\gamma$ denotes the scaled trap strength  with respect to the viscous drag coefficient $\gamma$.
The translational Wiener process $d\bm{B}$ and the rotational process $dW$ both have zero mean, with variances $\langle dB_i dB_j \rangle = \delta_{ij} dt$ (for $i,j \in {x,y}$) and $\langle dW^2 \rangle = dt$.
The corresponding diffusion coefficients  $D$ (translational) and $D_r$ (rotational) lead to the units of time and length in terms of 
the persistence  time $\tau_r = D_r^{-1}$ and length scale $\ell = \sqrt{D/D_r}$.
The dimensionless control parameters are the Péclet number $\Pe = v_0/\sqrt{DD_r}$ (activity), chirality $\Omega = \omega/D_r$, and trap strength $\beta = \mu/D_r$.
The  dimensionless position  and time will be  denoted by $\tilde{\br} = \br / \ell$ and $\tilde{t} = t / \tau_r$.

\medskip

\noindent
The harmonic trap guarantees an asymptotic steady state, which we analyze using both exact analytic methods and numerical simulations. We numerically integrate the dimensionless version of the Langevin equations using the Euler-Maruyama scheme, comparing the results with our analytic predictions.
The steady-state moments are derived by formulating the Fokker-Planck equation, applying a Laplace transform, and utilizing the resulting moment-generating equations.
Key findings are summarized below, and the full analytic derivations are provided in the 
Appendix-\ref{appendix-A}.

\begin{figure*}
\centering
\includegraphics[width=8.75cm]{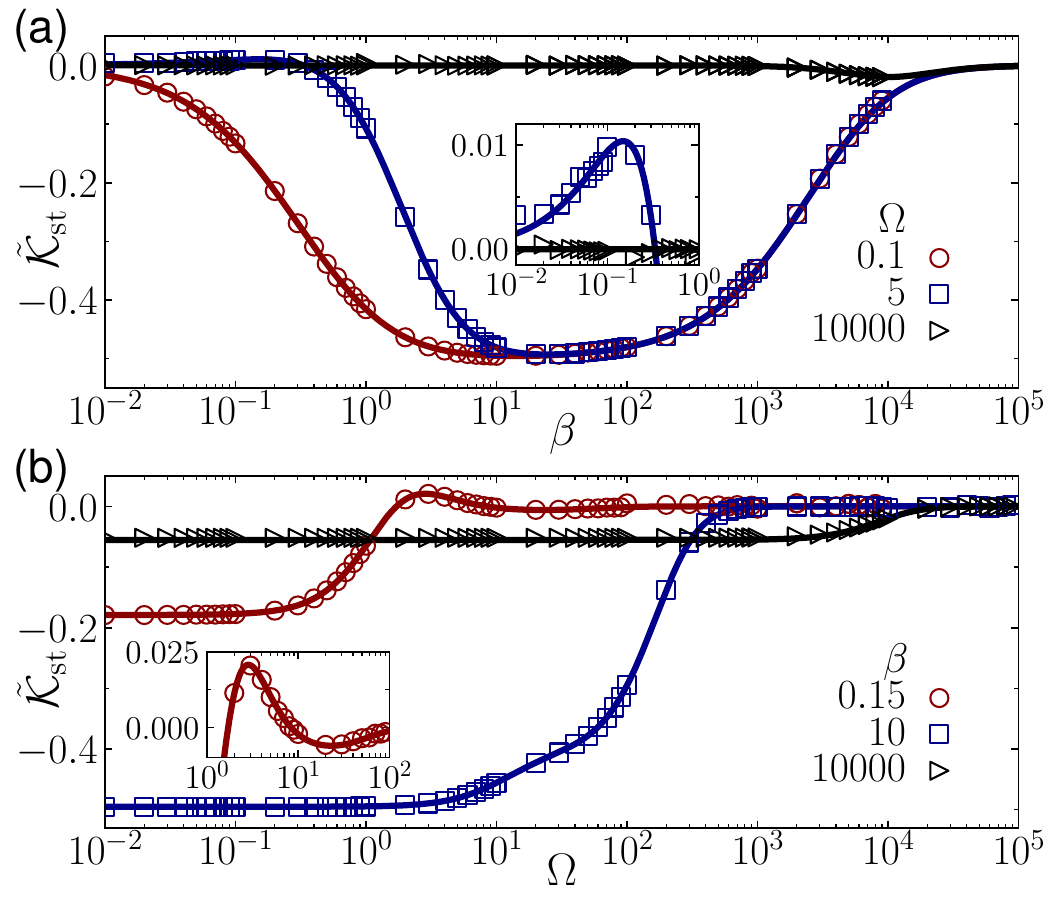}
\includegraphics[width=8.75cm]{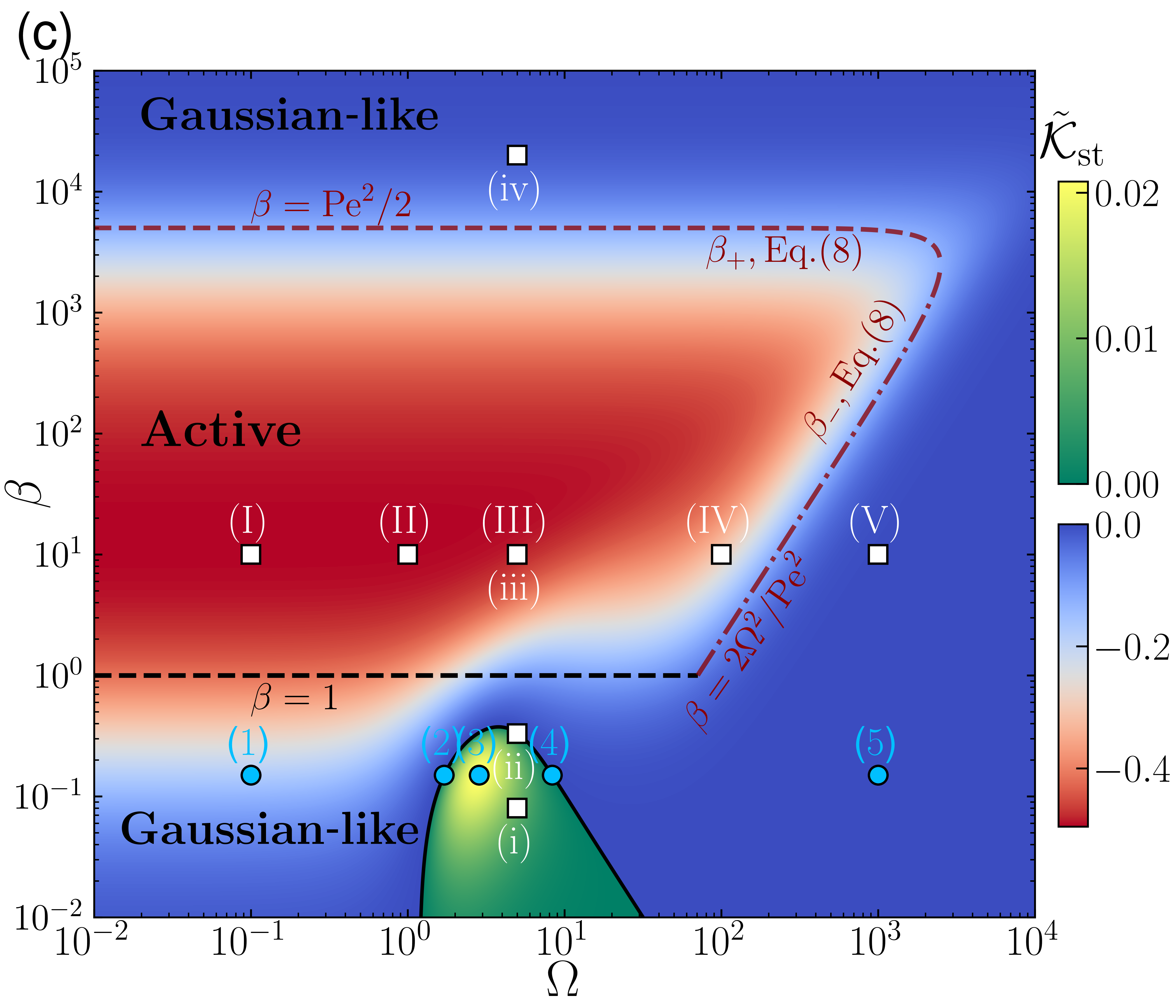}
\includegraphics[width=18cm]{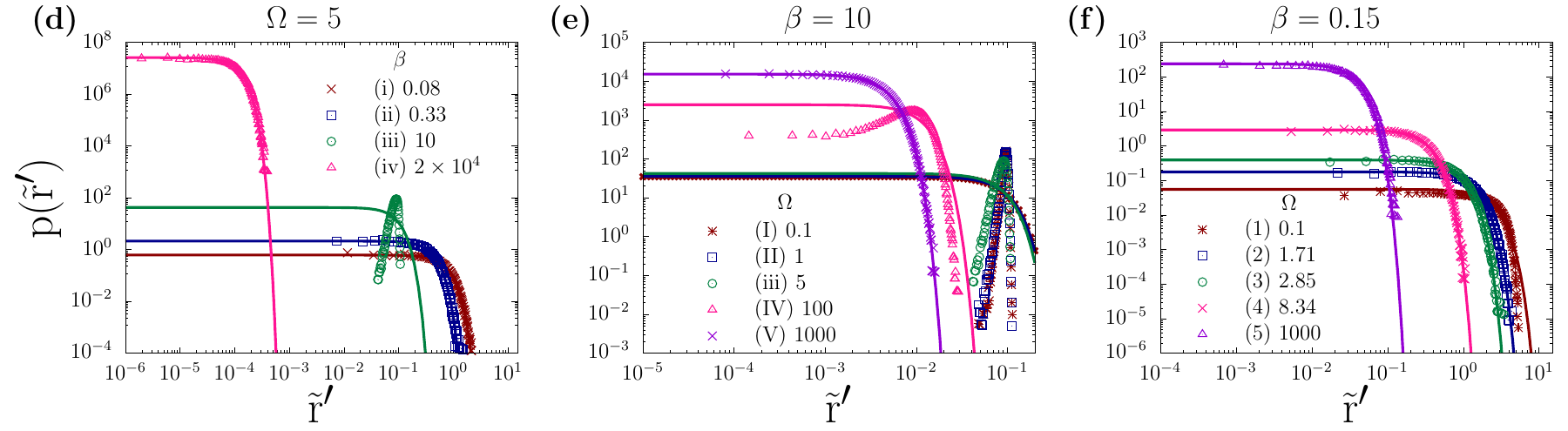}
\caption{
Steady state of a 2D chiral active Brownian particle (cABP) in a harmonic trap at $\Pe = 10^2$.
(a, b) Steady-state excess kurtosis $\tilde{\mathcal K}_{\rm st}$ as a function of (a) trap strength $\beta$ for $\Omega = 0.1$ ($\circ$), $5$ ($\square$), and $10^4$ ($\triangleright$); and (b) chirality $\Omega$ for $\beta = 0.15$ ($\circ$), $10$ ($\square$), and $10^4$ ($\triangleright$). Symbols: simulation; solid lines: analytical results.
(c) Phase diagram in the $(\Omega, \beta)$ plane, with the color map indicating $\tilde{\mathcal K}_{\rm st}$. 
Colored regions denote dynamical phases with bimodal (red), heavy-tailed (green), and Gaussian-like (blue) distributions.
Regions with $\tilde{\mathcal K}_{\rm st} > 0$ reflect weakly heavy-tailed active  behavior. The black contour ($\tilde{\mathcal K}_{\rm st} = 0$) marks the Gaussian limit. The dashed lines represent Eq.~\eqref{eq_beta} and its two asymptotic limits: $\beta = \Pe^2/2$ and $\beta = 2\Omega^2/\Pe^2$.
Numerical labels denote points where radial distributions are computed in (d–f).
(d–f) Radial probability distributions $\rm p(\tilde {\rm{ r}}')$ plotted versus $\tilde{\rm r}' = |\tbr|/\Pe$:
(d) Varying $\beta$ at fixed $\Omega = 5$ [points (i)–(iv) in (c)];
(e) Varying $\Omega$ at fixed $\beta = 10$ [points (I)–(V) and (iii)];
(f) Varying $\Omega$ at $\beta = 0.15$ [points (1)–(5)].
Symbols: simulation; solid lines: Gaussian reference. Off-center, non-Gaussian peaks at (I), (II), and (III)/(iii) correspond to $\tilde {\rm{ r}}' = \tilde {\rm{ r}}_{\rm ac}/\Pe$.
}
\label{fig2}
\end{figure*}

\subsection{Lower order moments at steady state}

The presence of the harmonic trap ensures a steady state with $\langle \tbr \ra_{\rm st}=0$. In equilibrium, the position distribution $\textrm {p}(\tbr)= \exp{\left(-{\tbr^2}/{2 \sigma^2}\right)}/{ \left(2 \pi \sigma^2\right)^{d/2} }$ is Gaussian, where $\sigma^2= \langle \tbr^2 \ra_{\rm st}/d$ in $d$-dimensions  with the mean-squared displacement (MSD) given by $\langle \tbr^2 \ra_{\rm st}=d/\beta$. This form is expected to hold at low enough activity. However, at large activity, even for non-chiral ABPs, the distribution deviates significantly from Gaussian behavior. These non-Gaussian departures can be  characterized by exact calculations of fourth moment and kurtosis of displacement~\cite{Chaudhuri2021}.
We compare the numerically obtained probability distributions with the Gaussian form supplemented by the exact MSD estimate shown below, to highlight the non-Gaussian departures (see solid lines in Figs.~\ref{fig2}(d,e,f)).  

In 2D, the distributions of non-chiral ABPs exhibited a circularly symmetric ring around the trap minimum, with particles climbing the trap potential until reaching a radial extent determined by the force balance condition: $\tilde {\rm{ r}}^0_{\rm ac}={\rm Pe}/\beta$~\cite{Malakar2020, Chaudhuri2021}. An equivalent estimate for chiral ABPs can be derived by analyzing the deterministic part of Langevin equations: $\dot {\tilde x} = {\rm Pe} \cos\phi - \beta \tilde x$, $\dot {\tilde y} = {\rm Pe} \sin\phi - \beta \tilde y$ and $\dot \phi = \Omega$. Solving these equations and obtaining the time-asymptotic expressions for $\tilde x(t)$ and $\tilde y (t)$, we find $\tilde {\rm{ r}}_{\rm ac}={\rm Pe}/\sqrt{\beta^2 + \Omega^2}$~(see Appendix-\ref{deterministic_soln_2d}). As we show later, this estimate agrees well with the off-center maxima of the probability distributions at high activity.

By applying the Fokker-Planck equation method 
{(detailed derivation presented in Appendix-\ref{appendix-A})}, the first result we obtain is the MSD at steady state,
\begin{align}
\langle \tbr^2\rangle_{\rm st}& =\frac{2}{\beta}+\frac{\Pe^2(1+\beta)}{\beta\left[(1+\beta)^2+\Omega ^2\right]}\,.
\label{eq:msd_2d_st}
\end{align}
The first term, denotes the equilibrium contribution in 2D, and the second term is entirely due to activity which increases as ${\rm Pe}^2$, but decreases quadratically with chirality as $\Omega^{-2}$. This result is consistent with observations in chiral active solid phases~\cite{Debets2023, Shee2024, Marconi2025}. 

\noindent
The expression allows us to obtain an estimate of an effective diffusion coefficient $D_{\rm eff} = \beta \la \tbr^2 \ra_{\rm st}/2$ gives
\begin{align}
D_{\rm eff} = 1+\frac{\Pe^2(1+\beta)}{2  \left[(\beta +1)^2+\Omega ^2\right]}\,.
\end{align}
In the limit of vanishing trap stiffness $\beta \to 0$, the expression reduces to the free particle result $D_{\rm eff} =1+ \Pe^2/2(1+\Omega^2)$~\cite{Pattanayak2024}. $D_{\rm eff}$ vanishes as $\beta^{-1}$ for large trap stiffness. 

\subsection{Steady-state excess kurtosis and phase diagram}
In the steady state, the deviation from Gaussian-like behavior is quantified by the excess kurtosis, defined as $\tilde{\cal{K}}_{\rm st} = {\langle{\tbr^4}\rangle_{\rm st}}/{\tilde{\mu}_4^{\rm st}} - 1$, with 
$\tilde{\mu}_4=\langle  \tbr^2\rangle^2 + 2\langle  \tilde{r}_i  \tilde{r}_j \rangle^2$ giving the estimate of the fourth moment in a Gaussian process. Although vanishing $\tilde{\cal{K}}_{\rm st}$ may in principle arise from non-Gaussian distributions via specific cancellations, our numerical simulations confirm that the $\tilde{\cal{K}}_{\rm st} \approx 0$ regimes identified here are indeed Gaussian-like. In 2D, $\tilde{\mu}_4^{\rm st} = 2 \langle \tbr^2 \rangle_{\rm st}^2$ gives the following exact expression (see  
Appendix-\ref{appendix-A})
\begin{widetext}
\begin{equation}
      \tilde{\cal{K}}_{\rm st}=\frac{\beta  \text{Pe}^4 \left(-2 (\beta +1) \left(5 \beta ^2+\beta -5\right) \Omega ^2-\beta  \Omega ^4-(\beta +1)^2 (\beta +2) (3 \beta +1) (3 \beta +7)\right)}{2 \left((\beta +2)^2+\Omega ^2\right) \left((3 \beta +1)^2+\Omega ^2\right) \left((\beta +1) \left(2 \beta +\text{Pe}^2+2\right)+2 \Omega ^2\right)^2} .
      \label{eq:exces_kurtosis_2d}
\end{equation}
\end{widetext}

Figs.~\ref{fig2}(a,b) display exact agreement of the analytic predictions (solid lines) with simulation results (symbols).
To account for the non-monotonic variation of the steady-state excess kurtosis observed in these figures, we analyze the limiting cases with respect to $\beta$ and $\Omega$ as detailed below.

Expanding around $\beta= 0$ we find,
\begin{align}
\tilde{\cal{K}}_{\rm st} &\simeq \frac{\beta  \text{Pe}^4 \left(5 \Omega ^2-7\right)}{\left(\Omega ^4+5 \Omega ^2+4\right) \left(\text{Pe}^2+2 \Omega ^2+2\right)^2} - \mathcal{O}(\beta^2)\, ,
\label{eq_K_small_beta}
\end{align}
with $\tilde{\cal{K}}_{\rm st}$ vanishing linearly with $\beta$.
In the small $\beta$ limit, $\tilde{\cal{K}}_{\rm st}$ behaves differently depending on $\Omega$: it is positive when $\Omega^2 > 7/5$ and negative when $\Omega^2 < 7/5$. As illustrated in Fig. \ref{fig2}(a), for instance, at $\Omega = 5$ (blue square symbols) where $\Omega^2 > 7/5$, $\tilde{\cal{K}}_{\rm st}$ is initially positive at small $\beta$, but then becomes negative, following a $-\beta^2$ dependence as the trap stiffness increases. After reaching a minimum at intermediate values of $\beta$, it begins to rise again, ultimately vanishing as $\sim -\beta^{-2}$ in the $\beta \to \infty$ limit.

We now examine the $\Omega$-dependence of $\tilde{\cal{K}}_{\rm st}$ in a similar way. Expanding around $\Omega= 0$ we find
\begin{align}
\tilde{\cal{K}}_{\rm st} &\simeq -\frac{\beta  \text{Pe}^4 (3\beta+7)}{2 (\beta+2)(3\beta+1) (2\beta+\Pe^2+2)^2}+\mathcal{O}(\Omega^2)\,.
\label{eq_K_small_omega}
\end{align}
It reveals that the zeroth-order term is independent of $\Omega$, while the next term, $+\mathcal{O}(\Omega^2)$, drives the negative excess kurtosis towards zero.
This behavior is depicted in Fig.~\ref{fig2}(b) across a range of trap stiffness values, from $\beta=0.15$ to $\beta=10^4$ traps.
As $\beta$ increases, the range of $\Omega$ with constant negative kurtosis broadens.
Under weak trapping, positive kurtosis surfaces at moderate chirality, while in the $\Omega \to \infty$ limit, $\tilde{\cal{K}}_{\rm st}$ vanishes as $-\Omega^{-4}$.

\medskip

\noindent
{\bf Phase diagram:} Fig.~\ref{fig2}(c) shows a {\it phase diagram} in the $\Omega-\beta$ plane, highlighting the transition between Gaussian (equilibrium-like) and non-Gaussian phases based on deviations from Gaussian statistics. While there is no true phase transition in this single-particle system, the diagram reveals a re-entrant crossover, shifting from Gaussian-like to non-Gaussian and back to Gaussian-like phases as trap stiffness increases.
We identify three phases based on steady-state excess kurtosis: negative kurtosis defines the active phase (red) with a symmetric bimodal distribution with off-center peaks; vanishing kurtosis corresponds to the phase (blue) with a Gaussian-like distribution; and small positive kurtosis indicates another active phase (green) with a weakly heavy-tailed distribution, unique to chiral ABPs and absent in trapped achiral ABPs~\cite{Chaudhuri2021}. The solid line represents $\tilde{\cal{K}}_{\rm st} = 0$, separating negative and positive kurtosis.

\medskip

\noindent
{\bf Probability distributions:} In Figs.~\ref{fig2}(d-f), we present radial position distributions for the parameter values marked by Roman and Indo-Arabic numerals in the phase diagram of Fig.~\ref{fig2}(c). The solid lines correspond to the Gaussian distribution 
$\rm p(\tbr)= \exp{\left(-{\tbr^2}/{\langle \tbr^2 \rangle_{\rm st}}\right)}/{ \pi \langle \tbr^2 \rangle_{\rm st} }$ 
with $\langle \tbr^2 \rangle_{\rm st}$ for the corresponding parameter values. 
These help identify deviations from equilibrium-like Gaussian behavior.

In Fig.~\ref{fig2}(d), the position distribution is shown for varying trap  stiffness $\beta=0.08~\mathrm{(i)}, 0.33~\mathrm{(ii)}, 10~\mathrm{(iii)}, 2\times 10^4~\mathrm{(iv)}$ with constant $\Omega=5$.
The transition is observed from a weakly heavy-tailed active distribution ($\tilde{\cal K}_{\rm st}>0$) to a Gaussian-like, then to an active state with off-center peak ($\tilde{\cal K}_{\rm st}<0$), and finally back to a Gaussian-like state.
Notably, for $\mathrm{(i)}~\beta = 0.08$, although the excess kurtosis is positive, it is small ($\tilde{\cal{K}}_{\rm st} \approx 0.0084$), and the distribution appears nearly Gaussian.  
For $\mathrm{(ii)}~\beta=0.33$ on the zero excess kurtosis line, a Gaussian distribution is expected and observed.
At $\mathrm{(iii)}~\beta=10$, a peak in the  distribution away from the trap center, at $\tilde {\rm{ r}}_{\rm ac}={\rm Pe}/\sqrt{\beta^2 + \Omega^2}$,  indicates an active state. At $\mathrm{(iv)}~\beta=2 \times 10^4$, the distribution reverts to Gaussian.

In Fig.~\ref{fig2}(e), radial distributions across different chirality values ($\Omega$) with $\beta=10$ are shown, revealing transitions from active (off-center peak at $\tilde {\rm{ r}}_{\rm ac}$) to Gaussian-like states.
Fig.~\ref{fig2}(f) at $\beta=0.15$ shows radial distributions at various $\Omega$ corresponding to small positive or negative kurtosis values with small departures from Gaussian.

\medskip

\noindent
{\bf Analysis of the phase diagram:} The phase diagram reveals an intermediate regime of $\beta$ supporting active dynamics marked by negative kurtosis. For $\beta > 1$, persistence dominates over trap relaxation time allowing for active bimodal phase. Neglecting diffusion, the maximal radial reach of a trapped particle is
$\tilde {\rm{ r}}_{\rm ac} = \Pe/\sqrt{\beta^2+\Omega^2}$~(see Appendix-\ref{deterministic_soln_2d}),
while translational noise broadens the distribution over $\ell_{\rm eq} = \sqrt{2/\beta}$.
This is the dimension-less form of $\ell_{\rm eq} \ell =\sqrt{2D/\mu}$.
The off-center peak becomes negligible when $\tilde {\rm{ r}}_{\rm ac} \lesssim \ell_{\rm eq}$, solution of the quadratic equation at $\tilde {\rm{ r}}_{\rm ac} = \ell_{\rm eq}$ defines two bounds of the active bimodal phase,
\begin{equation}
    \beta_{\pm} = \frac{\Pe^2}{4} \pm \left( \frac{\Pe^4}{16} - \Omega^2   \right)^{1/2}.
    \label{eq_beta}
\end{equation}
These boundaries are plotted in Fig.~\ref{fig2}(c). The active regime is confined to $\beta < {\Pe^2}/{4} + \left( {\Pe^4}/{16} - \Omega^2   \right)^{1/2}$ and $\beta > {\Pe^2}/{4} - \left( {\Pe^4}/{16} - \Omega^2   \right)^{1/2}$. The nature of these bounds becomes evident in the limiting cases: for large $\beta$, the inequality reduces to $\beta < \Pe^2/2$, while in the low-$\beta$ limit, it simplifies to $\beta > 2\Omega^2/\Pe^2$. The flattening of the lower boundary, is due to the requirement of $\beta > 1$.

The boundary of the weakly positive kurtosis observed in the low-$\beta$ active regime is not fully resolved, though, it may stem from a competition of the central thermal peak and activity-driven non-dominating probability weight at large radial distances. Nevertheless, the onset of the positive kurtosis  at small $\beta$  can be understood by considering Eq.\eqref{eq_K_small_beta} and the sign change occurring at $\Omega=\sqrt{7/5}$.       
%

\begin{figure*}
\centering
\includegraphics[width=8.5cm]{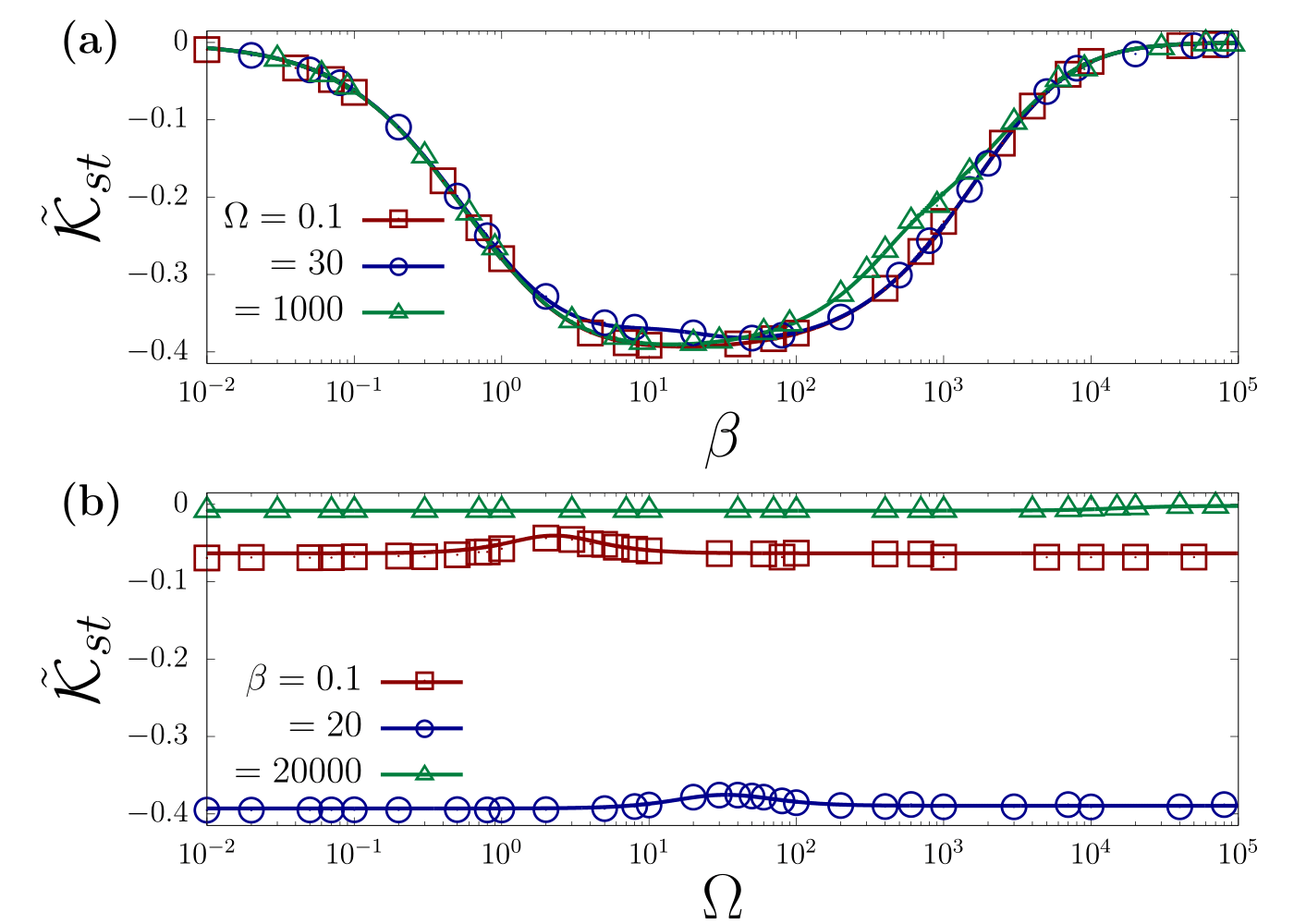}
\includegraphics[width=9.1cm]{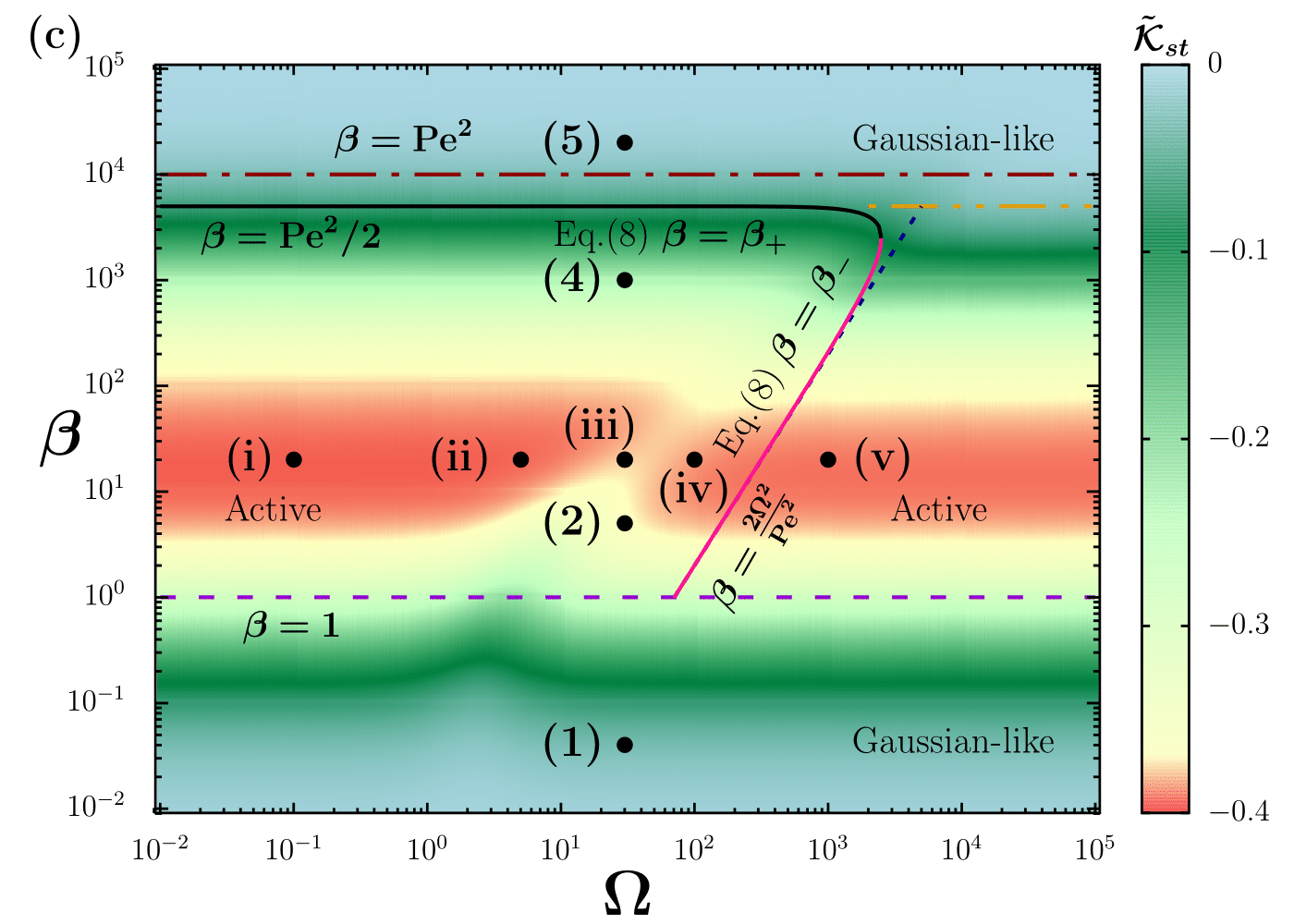}
\includegraphics[width=18cm]{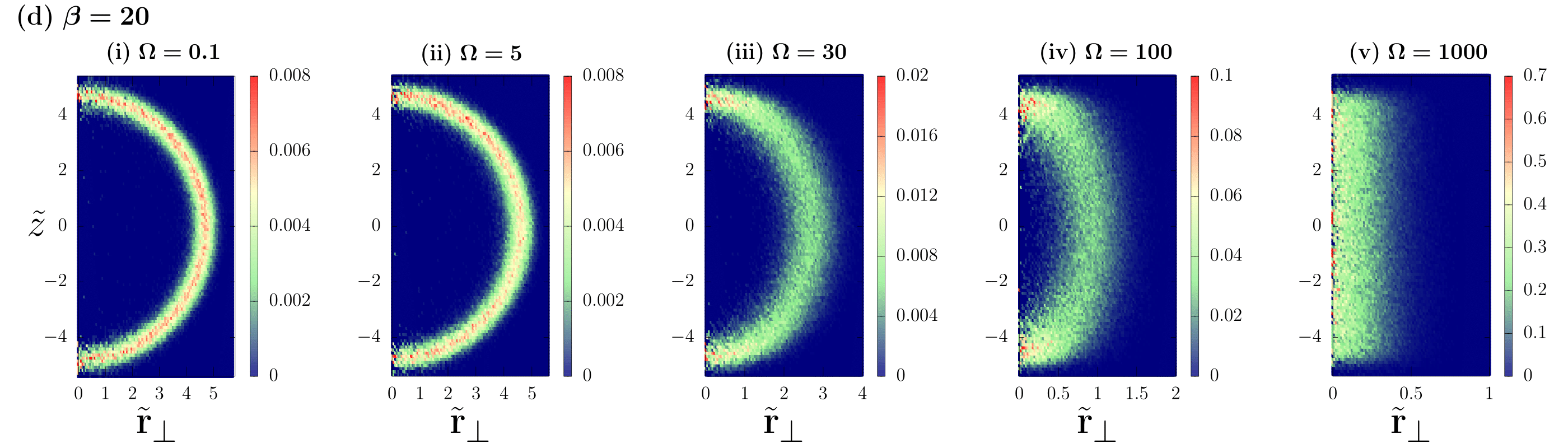}
\includegraphics[width=18cm]{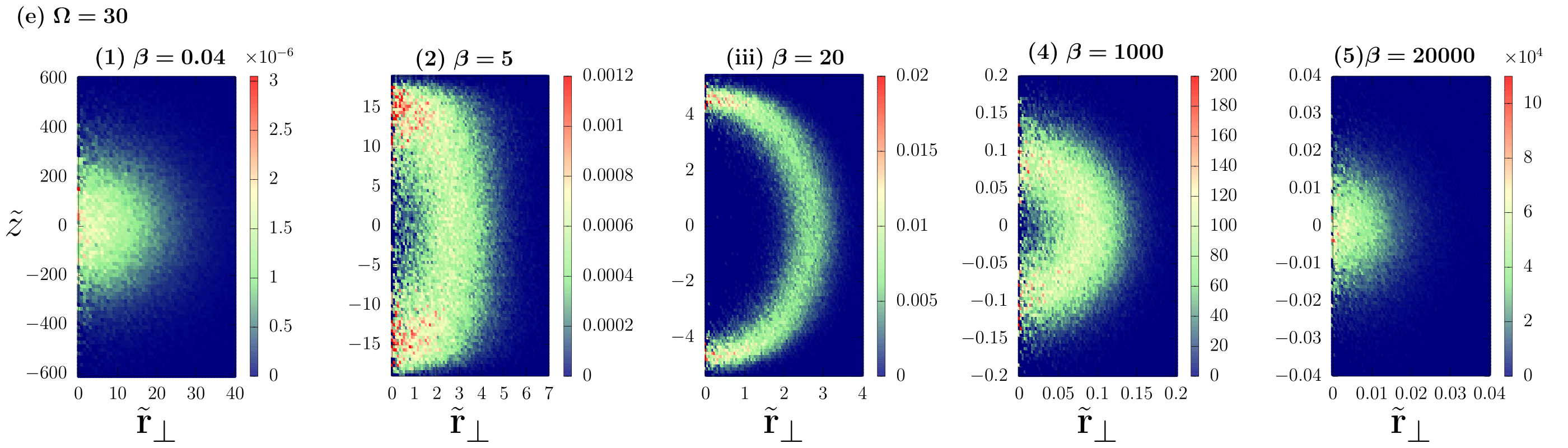}
\caption{Steady state of a 3D active Brownian particle (ABP) under torque in a harmonic trap at $\Pe = 10^2$.
(a, b) Steady-state excess kurtosis as a function of $\beta$ and $\Omega$. Symbols: simulation; solid lines: analytical predictions.
(c) Phase diagram in the $\Omega$–$\beta$ plane, with color indicating the steady-state excess kurtosis. The lines represent Eq.~\eqref{eq_beta} along with its two limiting cases: $\beta = \Pe^2/2$ and $\beta = 2\Omega^2/\Pe^2$. Also shown are the boundary values at $\beta = 1$ and $\beta = \Pe^2$.
(d, e) Probability distribution function plotted as a function of $\tilde{\rm r}_\perp$ and $\tilde{z}$.
In (d), $\Omega$ is varied at fixed $\beta = 20$, corresponding to points (i)–(v) in (c).
In (e), $\beta$ is varied at fixed $\Omega = 30$, corresponding to points (1), (2), (iii), (4), and (5) in (c).
}
\label{fig3}
\end{figure*}

%
\section{Trapped and torque-driven ABP in 3D} 
In 3D, the heading direction is given by $\bu = (\sin \theta \cos \phi, \sin \theta \sin \phi, \cos \theta)$, where $\theta$ and $\phi$ represent the polar and azimuthal angles,   respectively.
A torque  $\bm{\omega}= \omega_0 (\sin \theta_\omega \cos\phi_{\omega}, \sin\theta_\omega\sin\phi_\omega, \cos\theta_\omega)$ is considered, with amplitude $\omega_0$ and orientation defined by the angles $\theta_\omega$ and $\phi_\omega$ (see Fig~\ref{fig1}).
Evolution of the heading direction $\bu$ is controlled by the generalized force ${\bm\omega} \times {\bf \hat u}$ and orientational noise. 
Within the It\^o interpretation~\cite{Ito1975, VandenBerg1985, Raible2004, Mijatovic2020, Sevilla2016}, the Langevin dynamics can be expressed as:
\begin{align}
d{\br}(t) &= v_0\bu dt+\sqrt{2 D}\, d \bm{B}(t)-\mu \br dt\,,\label{eom1:3d}\\
d\theta(t) &= \omega_{\theta} dt+ \frac{D_r}{\tan\theta} dt+\sqrt{2 D_r}\, d W_\theta(t)\,,\label{eom2:3d}
\\
d\phi(t) &=  \omega_{\phi} dt+\frac{\sqrt{2 D_r}\, d W_\phi(t)}{\sin\theta}\,,
\label{eom3:3d}
\end{align}
where the generalized forces are $\omega_{\theta} = {\bf \hat u}_\theta \cdot ({\bm\omega} \times {\bf \hat u}) =  \omega_0 \sin\theta_\omega \sin(\phi_\omega-\phi)$ and
$\omega_{\phi}
= {\bf \hat u}_\phi \cdot ({\bm\omega} \times {\bf \hat u})=\omega_0\left(\cos\theta_\omega-\cot\theta\sin\theta_\omega \cos(\phi_\omega-\phi)\right)$~\footnote{The unit vectors ${\bf \hat u}_\theta = \partial \hat{\bm u}/\partial \theta$ and ${\bf \hat u}_\phi =(1/\sin\theta) (\partial \hat{\bm u}/\partial \phi)$.}. 
Considering a constant torque, ${\bm{\omega}} =  \omega_0 \hat{z}$ which implies $\theta_\omega=0$, we have:
\begin{align}
 d\theta(t) = \frac{D_r}{\tan\theta} dt+\sqrt{2 D_r}d W _\theta ; \,\,\, d\phi(t) =  \omega_0 dt+\frac{\sqrt{2 D_r}d W_\phi }{\sin\theta}.
 \end{align}
The translational and rotational Gaussian noise are characterized by $\langle d\bm{B}_{i}d\bm{B}_{j}\rangle=\delta_{ij}dt$, $\langle dW_\theta^2\rangle=dt$, $\langle dW_\phi^2\rangle=dt$ and $\langle d\bm{B}_{i}\rangle=\langle dW_\theta\rangle=\langle dW_\phi\rangle=0$.

\medskip

\noindent
It is straightforward to numerically integrate the above equations using the Euler-Maruyama scheme.
As before, we can formulate the Fokker-Planck equation, apply the Laplace transform, and perform the subsequent analysis to obtain the steady-state moments.
Below, we summarize the key findings, with detailed derivations 
presented in the 
Appendix-\ref{appendix-C}.

%
\subsection{Lower order moments at steady state}
It is important to note that the chosen torque induces rotation around the $z$-axis, resulting in motion confined to the $x-y$ plane, while dynamics along the $z$-axis follow a persistent random walk. This leads to dynamical asymmetry, which we address by separately computing the component of MSD in the $xy$-plane,
\begin{equation}
    \langle \tbr_\perp^2\rangle_{\rm st} = \frac{2}{\beta } + \frac{2\Pe^2 (2+\beta )}{3 \beta  \left((2 +\beta )^2+\Omega ^2\right)}\,,
\end{equation}
and that along the $z$-axis,
\begin{equation}
    \langle \tbr_\parallel^2\rangle_{\rm st} = \frac{1}{\beta } + \frac{\Pe^2}{3 \beta (2 +\beta )}\, ,
    \label{eq:MSD_st_parallel_3d}
\end{equation}
with the last expression independent of $\Omega$.
As expected, the dynamics along the $z$-axis agree with those of non-chiral ABPs in a harmonic trap~\cite{Chaudhuri2021}.
%
Thus the MSD describes a prolate spheroid with $\langle \tbr_\perp^2\rangle_{\rm st} < \langle \tbr_\parallel^2\rangle_{\rm st}$ for any finite $\Omega$, $\langle \tbr_\perp^2\rangle_{\rm st}$ shrinking with increasing $\Omega$. 

A symmetrized MSD in the steady state, $\la\tbr^2\ra_{\rm st}=\langle \tbr_\perp^2\rangle_{\rm st} + \langle \tbr_\parallel^2\rangle_{\rm st}$, reads 
\begin{align}
  \la\tbr^2\ra_{\rm st} &= \frac{3}{\beta} + \frac{\text{Pe}^2 \left(3 (\beta +2)^2+\Omega ^2\right)}{3 \beta[(\beta +2) \Omega ^2+(\beta +2)^3]} \, ,
  \label{eq:msd_3d_st_dimensionless}
\end{align}
where, the first term is due to translational noise, and the second term has entirely active origin. In the limit of vanishing $\beta$, it reduces to the known expression derived before~\cite{Pattanayak2024}.

\subsection{ Steady-state excess kurtosis and phase diagram}
We use the exact analytic expression for the steady-state excess kurtosis $\tilde {\cal{K}}_{\rm st}$ (provided in the 
Appendix-\ref{appendix-C}) to precisely quantify the non-Gaussian deviations. Figs.~\ref{fig3}(a) and \ref{fig3}(b) display $\tilde {\cal{K}}_{\rm st}$ as a function of $\beta$ and $\Omega$, respectively. The solid lines represent the analytic results, which show excellent agreement with the simulation results depicted by the points.

\medskip

\noindent
We plot $\tilde {\cal{K}}_{\rm st}$ as a function of $\beta$ for three $\Omega$ values $\Omega=0.1,~30,~10^3$ with $\Pe=10^2$ in Fig.~\ref{fig3}(a).
For an ABP under torque without a harmonic trap (or at small $\beta$ values), the long-time behavior shows zero excess kurtosis~\cite{Pattanayak2024}.
The steady-state excess kurtosis initially deviates towards negative values (indicating activity-dominated behavior) as trap strength increases, before eventually returning to Gaussian behavior (zero excess kurtosis).
This trend aligns with prior observations in two dimensions and with earlier studies of ABPs in a harmonic trap~\cite{Chaudhuri2021}.
In contrast to two dimensions, in three dimensions the steady state excess kurtosis remains negative, indicating activity-dominated behavior in the high chirality regime.
The behavior of the steady state excess kurtosis is qualitatively independent of chirality values because the persistence motion of particles is not restricted for high chirality, effectively displaying position distribution along the $z$-axis.
We plot $\tilde {\cal{K}}_{\rm st}$ as a function of $\Omega$ for three $\beta$ values $\beta=0.1,~20,~2\times 10^4$ with $\Pe=10^2$ in Fig.~\ref{fig3}(b).
For small $\beta$ values, the behavior of the particle is nearly Gaussian (excess kurtosis $\tilde {\cal{K}}_{\rm st}\sim 0$) and independent of chirality.
For $\beta=0.1$, the steady-state excess kurtosis $\tilde {\cal{K}}_{\rm st}$ shows negative small values, indicating low activity regimes.
For intermediate trap strength $\beta=20$, steady-state excess kurtosis $\tilde {\cal{K}}_{\rm st}$ shows large negative values, indicating highly active regimes.
The probability distribution of the position vector exhibits a bimodal form, with off-center peaks located at $\tilde {\rm{r}}^{\rm ac}_{\parallel}= \Pe/\beta$ along the $z-$axis and at $\tilde {\rm{r}}^{\rm ac}_{\perp}=\Pe/\sqrt{\beta^2+\Omega^2}$ in the $x-y$ plane.
For high value of trap strength $\beta=2\times 10^{4}$, steady-state excess kurtosis $\tilde {\cal{K}}_{\rm st}$ shows the Gaussian behavior (nearly zero excess kurtosis), independent of chirality $\Omega$.

\medskip

\noindent
Fig.~\ref{fig3}(c) shows the steady-state phase diagram in the $\Omega-\beta$ plane, highlighting different states observed using excess kurtosis.
The excess kurtosis $\tilde {\cal{K}}_{\rm st}$ quantify the bimodal active (negative excess kurtosis) and Gaussian-like (near-zero negligibly small negative excess kurtosis) region.
For intermediate values of trap strength $\beta$, steady-state excess kurtosis takes negative values around $\tilde {\cal{K}}_{\rm st}\sim -0.4$, known as the activity dominated regime.
And, for small and large values of $\beta$, $\tilde {\cal{K}}_{\rm st}$ exhibits near zero values of Gaussian-like regime.
The phase diagram exhibiting re-entrant transition with $\beta$, similar to the behavior of ABP in a harmonic trap as presented in~\cite{Chaudhuri2021}.
The excess kurtosis exhibits qualitatively similar behavior regardless of chirality values, unlike the two dimensional case where high chirality values leads to Gaussian-like behavior discussed previously (see Fig.~\ref{fig2}(c)).
Thus, analyzing the position probability distribution in the $\tilde {\rm{ r}}_\perp-\tilde{z}$ plane for varying chirality values $\Omega = 0.1,~5,~30,~10^2,~10^3$ at $\beta = 20$, as shown in (d), is crucial for understanding the spatial behavior.

\bigskip 

\noindent
{\bf Analysis of the Phase diagram}:  
In contrast to the 2D case, the 3D scenario features two distinct active length scales, $\tilde {\rm{ r}}_\parallel^{\rm ac} = \Pe/\beta$, $\tilde {\rm{ r}}_\perp^{\rm ac} = \Pe/\sqrt{\beta^2 + \Omega^2}$, defined along and perpendicular to the torque direction, respectively, adding complexity to the phase diagram. The active regime is bounded by three key conditions: (i)~$\beta > 1$, ensuring sufficient persistence relative to the trap relaxation time; (ii)~$(\tilde {\rm{ r}}_\parallel^{\rm ac})^2 > 1/\beta$, or equivalently $\beta < \Pe^2$, which guarantees that the off-center peak along the direction of torque exceeds thermal broadening; and (iii)~$(\tilde {\rm{ r}}_\perp^{\rm ac})^2 > 2/\beta$, corresponding to Eq.\eqref{eq_beta}, ensuring appreciable off-center peak  perpendicular to the torque. As shown in the analysis of Eq.~\eqref{eq_beta}, this final condition is well approximated by $\beta < \Pe^2/2$ and $\beta > 2\Omega^2/\Pe^2$.
The boundaries plotted in Fig.\ref{fig3}(d) capture the key features of the phase diagram shown in Fig.\ref{fig3}(c), which is derived from the full expression of excess kurtosis. The segment between points $(iii)$ and $(iv)$ in Fig.~\ref{fig3}(c) marks a transition from a fully active regime to one where only the dynamics along the $z$-axis remain active, while motion in the $xy$-plane becomes effectively Gaussian-like, characterized by trapped diffusion with an activity-dependent diffusivity. In 3D, $z$-dynamics remain active for  $1<\beta <\Pe^2$ at all $\Omega$, thereby permitting a re-entrant transition with respect to $\beta$ across all values of 
$\Omega$.
 
\medskip

\noindent
{\bf Probability distributions:} 
In Figs.~\ref{fig3}(d) and (e), we show the steady-state position distributions $p(\tilde {\rm {r}}_\perp, \tilde z)$, with $\tilde {\rm{ r}}_\perp = \sqrt{\tilde x^2 + \tilde y^2}$. Figure~\ref{fig3}(d) illustrates how the distribution evolves with increasing $\Omega$ at fixed $\beta = 20$. The large activity ($\Pe = 10^2$) keeps the system well within the active phase up to $\Omega \approx 10^3$, consistent with the bound $\Omega < \Pe^2/4$. Accordingly, all distributions remain in the active regime and exhibit pronounced non-Gaussian features.  At low $\Omega$, the profiles are approximately spherically symmetric; as $\Omega$ increases, the distribution becomes increasingly anisotropic, narrowing along $\tilde {\rm{ r}}_\perp$ and eventually forming a band-like shape. In contrast, the spread along $\tilde z$ remains largely unchanged.
This anisotropic deformation is governed by two distinct active length scales: $\tilde {\rm{ r}}_\parallel^{\rm ac} = \Pe/\beta$ (along $\tilde z$) and $\tilde {\rm{ r}}_\perp^{\rm ac} = \Pe/\sqrt{\beta^2 + \Omega^2}$ (in the transverse plane). Since $\tilde {\rm{ r}}_\parallel^{\rm ac}$ is independent of $\Omega$, the distribution width along $\tilde z$ remains fixed. In contrast, the decrease of $\tilde {\rm{ r}}_\perp^{\rm ac}$ with increasing $\Omega$ explains the shrinking spread in $\tilde {\rm{ r}}_\perp$. These analytical expressions closely match the peak positions and shapes observed in the simulations.

Figure~\ref{fig3}(e) shows the evolution of the position distribution at fixed activity $\Pe = 10^2$ and torque $\Omega = 30$, as the trap strength $\beta$ is varied. The distribution exhibits approximate spherical symmetry in both the small- and large-$\beta$ limits. At low $\beta = 0.04$, limited persistence results in equilibrium-like behavior with a Gaussian distribution. At the opposite extreme, for $\beta = 2 \times 10^4 > \Pe^2$, thermal fluctuations dominate the dynamics in all directions, and the distribution again becomes Gaussian.
In the intermediate regime, $5 \leq \beta \leq 10^3$, activity plays a leading role, giving rise to distinctly non-Gaussian, anisotropic distributions. These are well described by the active length scales $\tilde {\rm{ r}}_\parallel^{\rm ac} = \Pe/\beta$ and $\tilde {\rm{ r}}_\perp^{\rm ac} = \Pe/\sqrt{\beta^2 + \Omega^2}$. For smaller $\beta$ (e.g., $\beta = 5 < \Omega$), $\tilde {\rm{ r}}_\parallel^{\rm ac} > \tilde {\rm{ r}}_\perp^{\rm ac}$, producing a band-like distribution elongated along the $z$-axis. As $\beta$ increases, $\tilde {\rm{ r}}_\perp^{\rm ac}$ approaches $\tilde {\rm{ r}}_\parallel^{\rm ac}$, gradually restoring isotropy. This crossover manifests in a transition from band-like to ring-like distributions.

\section{Conclusions} 
Our work establishes a quantitative framework for understanding how chirality, activity, and confinement jointly sculpt the non-equilibrium steady states of active particles. By applying a Laplace-transform approach to the Fokker–Planck equation, we derived exact expressions for displacement moments and the excess kurtosis, providing a sensitive and quantitative probe of non-Gaussian statistics in trapped active systems.  

This framework reveals three characteristic regimes: negative kurtosis corresponding to bimodal, off-center active states; vanishing kurtosis indicating Gaussian-like behavior; and small positive kurtosis associated with weakly heavy-tailed distributions, a feature unique to two-dimensional chiral particles. The dimensionality of the system dictates to what extent re-entrant transitions survive. 
In two dimensions, enhanced chirality diminishes non-Gaussian, bimodal features, promoting a transition toward Gaussian-like distributions. In three dimensions, torque preserves the non-Gaussian phase by sustaining trapped, non-chiral–like active motion along the torque axis, leading to anisotropic steady states.
Simple active length-scale arguments capture the resulting phase boundaries and account for the observed shape of steady-state distributions.

Beyond their immediate relevance for trapped microswimmers, our predictions outline experimentally accessible signatures—such as kurtosis crossovers and anisotropic steady-state distributions — that can serve as benchmarks for artificial and biological realizations of chiral active matter. Our results are amenable to experimental verification in systems such as L-shaped colloidal swimmers~\cite{Kummel2013}, chiral granular rotors~\cite{Caprini2025SelfWrapping}, and air-driven spinners~\cite{Lopez2022}. More broadly, our analysis demonstrates how confinement can be exploited to reveal hidden structure in active fluctuations, suggesting routes to engineer controllable steady states in chiral and torque-driven active matter.

\section*{Acknowledgments}
AP and AC thanks the computing facility at IISER Mohali. AC acknowledges support from the Indo-German grant (IC-12025(22)/1/2023-ICD-DBT).  D.C. acknowledges the Department of Atomic Energy (India) for grant 1603/2/2020/IoP/R\&D-II/150288, CY Cergy Paris Universit{\'e} for a Visiting Professorship, and ICTS–TIFR, Bangalore, for an Associateship.

\section*{Conflict of Interest}
The authors declare no conflicts of interest.

\section*{Author contributions}
DC and AC conceived and designed the research. AP and AS performed the analytical and numerical calculations. DC developed the active length-scale argument used to rationalize the phase diagrams and shapes of probability distributions. All authors contributed equally to the preparation of the manuscript.

\section*{Data availability} 
All data that support the findings of this study are included within the article. 

\appendix

\begin{widetext}
\section*{Appendix}
We derive, step by step, all the steady-state moments discussed in the main text using an exact analytical framework based on the Laplace transform method—originally developed for semiflexible polymers~\cite{Hermans1952} and recently adapted to active systems~\cite{Shee2020, Chaudhuri2021, Pattanayak2024}.

\section{Derivation of analytic moments in two-dimension}
\label{appendix-A}
We explicitly show the derivation of steady-state moments for chiral active Brownian particles in a harmonic trap in two dimensions (2d).

\subsection{\bf Moments generator equation}

The probability distribution function $P(\textbf{r},\hat{\textbf{u}},t)$ of the particle, described in Eqs.~\eqref{eom1:2d} and \eqref{eom2:2d} in main text, follows the Fokker-Planck equation
\begin{align}
\partial_t P &=D \nabla^2 P+ D_r\partial_\phi^2 P+\nabla\cdot[(\mu  \br- v_0\hat{\textbf{u}}) P]-\omega\partial_\phi P\,.
\label{FPE_dimensionless_2d}
\end{align}
By performing a Laplace transform $\tilde{P}(\br,\bu,s) = \int_0^\infty dt e^{-st} P (\br,\bu,t)$, the Fokker-Planck equation can be expressed as
\begin{align}
-P(\textbf{r},\hat{\textbf{u}},0)+s\tilde{P}(\textbf{r},\hat{\textbf{u}},s)=D\nabla^2\tilde{P} +D_r\partial_\phi^2\tilde{P}+\mu \nabla\cdot(\br \tilde{P}) - v_0\bu\cdot\nabla\tilde{P} -\omega\partial_\phi \tilde{P}\,,
\label{LFPE2d}
\end{align}
where the initial condition at $t=0$ is set by $P(\br,\bu,0)=\delta(\br)\delta(\bu -\bu_0)$, without any loss of generality. Finally, this leads to the moments generator equation
\begin{align}
-\langle\psi\rangle_0+s\langle\psi\rangle_s=v_0\langle\hat{\textbf{u}}\cdot\nabla\psi\rangle_s-\mu \langle \br \cdot\nabla\psi\rangle_s+D\langle\nabla^2\psi\rangle_s+D_r\langle\partial_\phi^2\psi\rangle_s+\omega\langle\partial_\phi\psi\rangle_s~,
\label{ME2}
\end{align}
for the mean of an arbitrary dynamical variable $\psi$ defined as $\langle\psi\rangle_s = \int d\br d\bu \psi(\br,\bu)\tilde{P}(\br,\bu,s)$, where the initial condition $\langle \psi \rangle_0 = \int d\br d\bu\, \psi(\br,\bu)\, P(\br,\bu,0)$. 

We use Eq.~\eqref{ME2} to derive the dynamical moments and then characterize the steady-state properties in the long time limit using them, which is the main focus here. To compute the exact expressions of different moments in the long-time limit or in the steady state, we use the final value theorem that says, 
\begin{align}
    \lim_{t \to \infty} f(t)= \lim_{s\to 0+} s \tilde{f}(s)\,,
    \label{eq:final_value}
\end{align}
where $\tilde{f}(s)= \int_0^\infty dt e^{-st} f(t)$ is the Laplace transform of the function $f(t)$.

\subsection{\bf Mean squared displacement (MSD)} 
\label{appendix_MSD_2d}
To calculate mean squared displacement, we consider $\psi=\br^2$ in Eq.~\eqref{ME2} gives 

\begin{align}
\langle \br^2\rangle_s=\frac{1}{(s+2\mu)}\left[4D\langle 1\rangle_s+2v_0\langle \br\cdot\bu\rangle_s\right]\,.
\label{eq:r2avg_Laplace_2d}
\end{align}
Putting $\psi=\br.\bu$ in the moment generating equation, Eq.~\eqref{ME2} we obtain
\begin{align}
\langle \textbf{r}\cdot\hat{\textbf{u}}\rangle_s=\frac{v_0(s+D_r+\mu)}{s[(s+D_r+\mu)^2+\omega^2]}\,. 
\label{eq:ur_cross_corr_Laplace_2d}
\end{align}
Using $\la1\ra_s=1/s$ and substituting Eq.~\eqref{eq:ur_cross_corr_Laplace_2d} into Eq.~\eqref{eq:r2avg_Laplace_2d}, we get the Laplace transformed expression of the mean squared displacement.
\begin{align}
    \langle \br^2\rangle_s=\frac{1}{(s+2\mu)}\left[\frac{4D}{s}+2v_0^2\frac{(s+D_r+\mu)}{s[(s+D_r+\mu)^2+\omega^2]} 
    \right]\,.
\label{eq:r2avg_Laplace_2d_final}
\end{align}
Applying the final value theorem, we obtain the MSD in the long-time limit $\la \br^2\ra_{\rm st}= \lim_{t\to\infty}\la \br^2\ra(t) = \lim_{s\to 0+}s \la \br^2\ra_s$,
\begin{align}
   \la \br^2\ra_{\rm st}=\frac{2 D}{\mu}+ \frac{v_0^2(D_r+\mu)}{\mu  \left(\omega ^2+(D_r+\mu)^2 \right)}\,.
\label{eq:msd_2d_st_dimensional}
\end{align}
In the dimensionless scale, it takes the form~\cite{Debets2023, Shee2024}, presented in main text in Eq.~\eqref{eq:msd_2d_st}.

\subsection{\bf Fourth moment of displacement}
%
Putting $\psi=\br^4$ into moments generator Eq.~\eqref{ME2}, we get
\begin{align}
\left\langle \br^{4}\right\rangle_s=\frac{1}{(s+4 \mu)}\left[16 D\left\langle \br^2\right\rangle_s+4 v_0\left\langle \br^2 \br\cdot \bu\right\rangle_{s}\right]\,.
\label{r4moment}
\end{align}
The second term can be calculated following the steps given below,

\medskip

\noindent
{\bf{Calculation of $\left\langle \br^2 \br\cdot \bu\right\rangle_{s} $}}

\begin{align}
&\left[s+D_r+3\mu+\frac{\omega^{2}}{s+D_r+3\mu}\right]\langle \br^2 \br\cdot \bu\rangle_s =8 D\left[\langle\br \cdot \bu\rangle_{s}+\frac{\omega\langle \partial_\phi(\br\cdot \bu)\rangle_{s}}{s+D_r+3\mu}\right]
+ v_0\left[2 \left\langle(\br \cdot \bu)^{2}\right\rangle_{s} 
+\frac{\omega\langle \partial_\phi(\br \cdot \bu)^{2}\rangle_{s}}{s+D_r+3\mu}+\langle \br^2\rangle_{s}\right]\,.
\end{align}

\noindent
{\bf{Calculation of $\left\langle(\br \cdot \bu)^{2}\right\rangle_{s}$}}

\begin{align}
&\left[s+4 D_r+2\mu+\frac{4 \omega^{2}}{s+4 D_r+2\mu}\right]\left\langle(\br\cdot \bu)^{2}\right\rangle_{s}= \frac{2 D}{s}+2 D_r\left\langle \br^2\right\rangle_{s}+2 v_0\langle\br\cdot \bu\rangle_{s} +\frac{2 \omega v_0\left\langle \partial_\phi(\br\cdot \bu)\right\rangle_{s}}{s+4 D_r +2\mu} +2 \omega^{2} \frac{\left\langle \br^2\right\rangle_{s}}{s+4 D_r +2\mu}\,. 
\end{align}
Again,
\begin{align}
\left\langle \partial_\phi(\br\cdot \bu)^{2}\right\rangle_{s}  &= \frac{1}{\left(s+4 D_r+2 \mu\right)}\left[2 v_0\left\langle \partial_\phi(\br\cdot \bu)\right\rangle_{s}-4 \omega\left\langle(\br\cdot \bu)^{2}\right\rangle_{z} +2 \omega\left\langle \br^2\right\rangle_{s}\right]\,.
\end{align}
The dimensionless expression of the fourth order moment of displacement in the steady state $\langle \tbr^4\rangle_{\rm st}=\lim_{s\to 0+}s \langle \tbr^4\rangle_s$,
\begin{align}
\langle \tbr^4\rangle_{\rm st}
& =\frac{8}{\beta ^2}+\frac{8 (\beta +1) \text{Pe}^2}{\beta ^2 \left((\beta +1)^2+\Omega ^2\right)}+\frac{\text{Pe}^4 \left(9 \beta ^4+42 \beta ^3+67 \beta ^2+(\beta  (\beta +4)+2) \Omega ^2+42 \beta +8\right)}{\beta ^2 \left((\beta +1)^2+\Omega ^2\right) \left((\beta +2)^2+\Omega ^2\right) \left((3 \beta +1)^2+\Omega ^2\right)}\,.
\end{align}

\subsection{\bf Excess kurtosis}
The deviation from the Gaussian process is measured by the excess kurtosis which can be written as,
\begin{equation}
    \tilde{\cal{K}} = \frac{\langle{\tbr^4}\rangle}{\tilde \mu_4} - 1\,,
    \label{ex_kurt}
\end{equation}
where 
\begin{equation}
\tilde{\mu}_4=\langle \delta \tbr^2\rangle^2 + 2\langle \delta \tilde{r}_i \delta \tilde{r}_j \rangle^2 + 2\langle \delta \tbr^2\rangle\langle{\tbr}\rangle^2
+ 4\langle \tilde{r}_i\rangle\langle \tilde{r}_j \rangle\langle \delta \tilde{r}_i\delta \tilde{r}_j \rangle + \langle {\tbr}\rangle ^4.
\nonumber
\end{equation}
This is a result, valid for any $d$-dimensional Gaussian process. 
In the presence of harmonic trap with the minimum at $\tbr =0$, the mean particle position $\langle \tbr \rangle=0$, and thus $\mu_4$ reduces to,
\begin{align}
    \tilde{\mu}_4=\langle  \tbr^2\rangle^2 + 2\langle  \tilde{r}_i  \tilde{r}_j \rangle^2. 
    \label{mu4_avg}
\end{align}
This result will be used for both two and three dimensional systems considered in this paper.
In two dimensions, the chiral ABP follows $\langle \tilde{r}_i \tilde{r}_j \rangle=\delta_{ij}\langle  \tbr^2\rangle/2$, and results in  
\begin{align}
    \tilde{\mu}_4=2 \langle  \tbr^2\rangle^2\,. 
    \label{mu4_avg_2d}
\end{align}
Thus, using the expressions of $\langle{\tbr^4}\rangle_{\rm st}$ and $\langle  \tbr^2\rangle_{\rm st}$, we get the exact expression of the steady-state excess kurtosis, presented in main text in Eq.~\eqref{eq:exces_kurtosis_2d}.

\medskip

\noindent
{\bf Limiting cases of steady state excess kurtosis}\\
Here, we list the limiting forms of the exact steady state excess kurtosis in two dimensions(Eq.~\eqref{eq:exces_kurtosis_2d} of the main text).
Weak trapping limit($\beta\to 0$) leads to
\begin{align}
&\lim_{\beta \to 0} \tilde{\cal{K}}_{\rm st} =\frac{\beta  \text{Pe}^4 \left(5 \Omega ^2-7\right)}{\left(\Omega ^4+5 \Omega ^2+4\right) \left(\text{Pe}^2+2 \Omega ^2+2\right)^2}
\nonumber\\ &-\beta^2\frac{\text{Pe}^4 \left(\text{Pe}^2 \left(\Omega ^8+17 \Omega ^6+219 \Omega ^4+463 \Omega ^2-172\right)+2 \left(\Omega ^2+1\right) \left(\Omega ^8-3 \Omega ^6+187 \Omega ^4+627 \Omega ^2-284\right)\right)}{2 \left(\Omega ^4+5 \Omega ^2+4\right)^2 \left(\text{Pe}^2+2 \Omega ^2+2\right)^3}+\mathcal{O}(\beta^3)~.
\label{eq:excess_kurtosis_2d_low_trapping}
\end{align}
This is correspond to Eq.~\eqref{eq_K_small_beta} in the main text. Strong trapping limit($\beta\to \infty$) gives
\begin{align}
&\lim_{\beta \to \infty} \tilde{\cal{K}}_{\rm st}=-\frac{\text{Pe}^4}{8 \beta ^2}+\frac{\text{Pe}^2\left(2+\text{Pe}^2\right)}{8 \beta^3}-\frac{\text{Pe}^4\left( 28+36 \text{Pe}^2+9 \text{Pe}^4-24 \Omega^2\right)}{96 \beta ^4}+O\left(\frac{1}{\beta ^5}\right)\,.
\end{align}
Low chirality limit($\Omega\to 0$) leads to
\begin{align}
&\lim_{\Omega \to 0} \tilde{\cal{K}}_{\rm st} = -\frac{\beta  (3 \beta +7) \text{Pe}^4}{2 \left(\left(3 \beta ^2+7 \beta +2\right) \left(2 \beta +\text{Pe}^2+2\right)^2\right)}+O\left(\Omega ^2\right)~.
\end{align}
This is correspond to Eq.~\eqref{eq_K_small_omega} in main text.
High chirality limit($\Omega\to \infty$) gives
\begin{align}
&\lim_{\Omega \to \infty} \tilde{\cal{K}}_{\rm st} =-\frac{\beta ^2 \text{Pe}^4}{8 \Omega ^4}+O\left(\frac{1}{\Omega ^6}\right)\,.
\end{align}
Low activity limit($\Pe\to 0$) leads to
\begin{align}
\lim_{\text{Pe} \to 0} \tilde{\cal{K}}_{\rm st} &= \text{Pe}^4\frac{\beta  \left(-2 (\beta +1) \left(5 \beta ^2+\beta -5\right) \Omega ^2-\beta  \Omega ^4-(\beta +1)^2 (\beta +2) (3 \beta +1) (3 \beta +7)\right)}{8 \left((\beta +1)^2+\Omega ^2\right)^2 \left((\beta +2)^2+\Omega ^2\right) \left((3 \beta +1)^2+\Omega ^2\right)}+O\left(\text{Pe}^6\right)\,.
\end{align}
High activity limit($\Pe\to \infty$) gives
\begin{align}
&\lim_{\text{Pe} \to \infty} \tilde{\cal{K}}_{\rm st} = \frac{\beta  \left(-2 (\beta +1) \left(5 \beta ^2+\beta -5\right) \Omega ^2-\beta  \Omega ^4-(\beta +1)^2 (\beta +2) (3 \beta +1) (3 \beta +7)\right)}{2 (\beta +1)^2 \left((\beta +2)^2+\Omega ^2\right) \left((3 \beta +1)^2+\Omega ^2\right)}+
\nonumber\\&\frac{2 \beta  \left((\beta +1)^2+\Omega ^2\right) \left(2 (\beta +1) \left(5 \beta ^2+\beta -5\right) \Omega ^2+\beta  \Omega ^4+(\beta +1)^2 (\beta +2) (3 \beta +1) (3 \beta +7)\right)}{(\beta +1)^3 \text{Pe}^2 \left((\beta +2)^2+\Omega ^2\right) \left((3 \beta +1)^2+\Omega ^2\right)}
      +O\left(\frac{1}{\Pe^4}\right)\,.
\end{align}
We use the above equations to analyze the exact steady state excess kurtosis behavior in the main text.

\section{\bf Deterministic solution of the particle trajectory in two-dimensions}
\label{deterministic_soln_2d}
In dimensionless scale, the deterministic parts of the Langevin equations, given in Eqs.~\eqref{eom1:2d} and \eqref{eom2:2d} of the main text, can be written as,
\begin{align}
     \dot {\tilde x} &= {\rm Pe} \cos\phi - \beta \tilde x\,,
     &&
     \dot {\tilde y} = {\rm Pe} \sin\phi - \beta \tilde y\,,
     &&
     \dot{\phi} = \Omega\,.
     \label{eq:deterministic_2d}
\end{align}

Integrating Eq~\eqref{eq:deterministic_2d} with respect to time and considering long-time limit, we get
\begin{align}
    \phi(\ttt) &= \Omega \ttt+ \phi_0\,,
    \\
    \nonumber
    \tilde{x}(\ttt) &= \cancelto{0}{\tilde x_0e^{-\beta \ttt}}+ \Pe\int_{-\infty}^{t}d \ttt^\prime e^{-\beta(\ttt-\ttt^\prime)}\cos \phi(\ttt^\prime)
    \\&= \Pe\int_{-\infty}^{t}d \ttt^\prime e^{-\beta(\ttt-\ttt^\prime)}\cos (\Omega\ttt^\prime+\phi_0)
    =\frac{\Pe}{\beta^2+\Omega^2}\left(\beta\cos(\Omega \ttt+\phi_0)+\Omega \sin(\Omega \ttt+\phi_0)\right)\,,
    \\
    \tilde{y}(\ttt) &= \cancelto{0}{\tilde y_0e^{-\beta \ttt}}+ \Pe\int_{-\infty}^{t}d \ttt^\prime e^{-\beta(\ttt-\ttt^\prime)}\sin \phi(\ttt^\prime)
     =\frac{\Pe}{\beta^2+\Omega^2}\left(\beta\sin(\Omega \ttt+\phi_0)-\Omega \cos(\Omega \ttt+\phi_0)\right)\,.
\end{align}
The trajectory in the long-time limit satisfies 
\begin{align}
\tilde{\rm r}^2(\ttt)=  \tilde{x}^2(\ttt) + \tilde{y}^2(\ttt)= \frac{\Pe^2}{\beta^2+\Omega^2}\,,
\end{align}
which is an equation of a circle of radius $\tilde{\rm r} _{\rm ac}= \Pe/\sqrt{\beta^2+\Omega^2}$. Equating $\tilde{\rm r} _{\rm ac}$ with the equilibrium spread $\ell_{\rm eq}=\sqrt{2/\beta}$ we obtain an approximate boundary that seperates the non-Gausssian regime with large negative excess kurtosis from the Gaussian regime with near-zero excess kurtosis(See Fig.~\ref{fig2}(c) and Fig.~\ref{fig3}(c)). The non-Gaussian probability distribution functions, obtained from simulations, have peaks close to $\tilde{\rm r} _{\rm ac}$ for two-dimensions(Figs~\ref{fig2}[d-f]). In three-dimensions, $\tilde{\rm r} _{\rm ac}$ provides a good estimate of peak-positions in the plane normal to the direction of the torque(Figs~\ref{fig3}[d-e]).

\section{Derivation of analytic moments in three-dimension}
\label{appendix-C}

We explicitly show the derivation of steady-state moments for active Brownian particles under torque in a harmonic trap in three dimensions (3d).

\subsection{\bf Moments generator equation}
The Fokker-Planck equation satisfied by the single particle probability distribution function, $P(\textbf{r},\hat{\textbf{u}},t)$,
\begin{align}
\partial_t P &=D\nabla^2 P+D_r\partial_\phi^2 P+\nabla\cdot[(\mu  \br- v_0\hat{\textbf{u}}) P]-\bm{\omega}\cdot\bR P\,.
\label{FPE_main}
\end{align}
Here $\bR\equiv \hat{\bm{u}}\times\nabla_\bu$ is equivalent to the gradient operator in the orientation vector space and $\mu=k/\gamma$. In the following sections, we have derived various moments associated with the motion of the particle.
By applying the Laplace transform to the Fokker-Planck equation,
\begin{align}
-P(\textbf{r},\hat{\textbf{u}},0)+s\tilde{P}(\textbf{r},\hat{\textbf{u}},s)=D\nabla^2\tilde{P} +D_r\bR^2\tilde{P}+\mu\nabla\cdot(\br \tilde{P}) - v_0\bu\cdot\nabla\tilde{P} -\bm{\omega}\cdot\bR \tilde{P}\,,
\label{LFPE3d}
\end{align}
where $\tilde{P}(\textbf{r},\hat{\textbf{u}},s)=\int_0^\infty dt e^{-st}P(\textbf{r},\hat{\textbf{u}},t)$ is the Laplace transformation of $P(\textbf{r},\hat{\textbf{u}},t)$ and $P(\textbf{r},\hat{\textbf{u}},0)=\delta^d(\textbf{r})\delta(\hat{\textbf{u}}-\hat{\textbf{u}}_0)$ is the initial probability distribution function.
Let us consider an arbitrary function $\psi=\psi(\textbf{r},\hat{\textbf{u}})$. If we multiply equation~\eqref{LFPE3d} with $\psi$ and integrate w.r.t $\textbf{r}$ and $\hat{\textbf{u}}$ over the whole space,
\begin{align}
-\psi_0+s\langle\psi\rangle_s=D\langle\nabla^2\psi\rangle_s+D_r\langle\bR^2\psi\rangle_s+v_0\langle\hat{\textbf{u}}\cdot\nabla\psi\rangle_s-\mu \langle \br \cdot\nabla\psi\rangle_s+\langle\boldsymbol{\Omega}\cdot\bR\psi\rangle_s\,,
\end{align}
where $\psi_0=\int d \textbf{r}\int d \hat{\textbf{u}} P(\textbf{r},\hat{\textbf{u}},0) \psi(\textbf{r},\hat{\textbf{u}})$ and $\langle\psi\rangle_s=\int d \textbf{r}\int d \hat{\textbf{u}} \tilde{P}(\textbf{r},\hat{\textbf{u}},s) \psi(\textbf{r},\hat{\textbf{u}})$
Let us consider the direction of the constant torque along the $z-$ axis, i.e. $\bm{\omega}=\omega \hat{z}$, which leads to a helical trajectory with a circular motion in the $x-y$ plane. The above equation for computing the moments further simplifies to,
\begin{align}
-\psi_0 +s\langle\psi\rangle_s =D\langle\nabla^2\psi\rangle_s+D_r\langle\bR^2\psi\rangle_s+v_0\langle\hat{\textbf{u}}\cdot\nabla\psi\rangle_s-\mu \langle \br \cdot\nabla\psi\rangle_s+\omega\langle\bR_z\psi\rangle_s\,.
\label{ME3}
\end{align}
Using the above equation, we compute the Laplace-transformed moments as outlined below. The main focus is on the particle's steady-state behavior discussed in the main text. The steady-state expressions of required dynamical moments are calculated in this article using the final value theorem given by Eq.~\eqref{eq:final_value}.

\subsection{\bf Mean squared displacement (MSD)} 
To calculate mean squared displacement (MSD), we consider $\psi=\br^2$ in Eq.~\eqref{ME3}, which leads to 
\begin{align}
\langle \br^2\rangle_s=\frac{1}{s+2\mu}\left[6D\langle 1\rangle_s+2v_0\langle \textbf{r}\cdot\hat{\textbf{u}}\rangle_s\right]\,.
\label{eq:r2avg_Laplace_3d}
\end{align}
where $\langle 1\rangle_s=1/s$. Setting $\psi=\tbr\cdot\bu$ in Eq.~\eqref{ME3} we get, 
\begin{align}
\langle \textbf{r}\cdot\hat{\textbf{u}}\rangle_s=\frac{v_0(s+2 D_r+\mu)+ \omega^2 D_r\langle z\text{u}_z\rangle_s}{\left[(s+2 D_r+\mu)^2+\omega^2\right]}\,.
\label{eq:ru_cross_corr_Laplace_3d_final}
\end{align}

Following the same procedure, we calculate $\langle z\text{u}_z\rangle_s$, 
\begin{align}
&\langle z\text{u}_z\rangle_s=\frac{v_0(2 D_r/s+\text{u}_{z0}^2)}{(s+2 D_r+\mu)(s+6 D_r)}\,,
\label{eq:zuz}
\end{align}
Substituting Eq.~\eqref{eq:ru_cross_corr_Laplace_3d_final} and Eq.~\eqref{eq:zuz} into Eq.~\eqref{eq:r2avg_Laplace_3d} and using the final value theorem, we obtain the MSD in the steady state $\la\tbr^2\ra_{\rm st} = \lim_{t\to \infty} \langle \tbr^2\rangle(t) = \lim_{s\to 0+} s\langle \tbr^2\rangle_s$. In dimensionless form, the steady state MSD presented in the main text in Eq.~\eqref{eq:msd_3d_st_dimensionless}.

\subsection{\bf Fourth moment of displacement}
Considering $\psi=\br^4$ into Eq.~\eqref{ME3} ,
\begin{align}
(s+4 \mu)\left\langle \br^{4}\right\rangle_s=20 D \left\langle \br^2\right\rangle_s+4 v_0\left\langle \br^2 \br\cdot \bu\right\rangle_{s}\,.
\label{r4moment3d}
\end{align}
Expression of $\left\langle \br^2\right\rangle_s$ is taken from the previous section. The second term can be calculated following the steps given below. In the course of finding some expressions we have encountered $R_z \psi$ which can be found following the procedure we did in the previous section for finding $\langle \textbf{r}\cdot\hat{\textbf{u}}\rangle_s$,

\medskip

\noindent
{\bf{Calculation of $\left\langle \br^2 \br\cdot \bu\right\rangle_{s}$}}

Choosing $\psi=\br^2 \br\cdot \bu$,
\begin{align}
&\left[s+2 D_r+3\mu+\frac{\omega^{2}}{s+2 D_r+3\mu}\right]\langle \br^2 \br\cdot \bu\rangle_s =10 D \left[\langle\br \cdot \bu\rangle_{s}+\frac{\omega\langle R_{z}(\br\cdot \bu)\rangle_{s}}{s+2 D_r+3\mu}\right]
\nonumber\\&+ v_0\left[2 \left\langle(\br \cdot \bu)^{2}\right\rangle_{s} 
+\frac{\omega\langle R_{z}(\br \cdot \bu)^{2}\rangle_{s}}{s+2 D_r+3 \mu}+\langle \br^2\rangle_{s}\right]+\frac{\omega^{2}\langle \br^2 z \text{u}_{z}\rangle_{s}}{s+2 D_r+3\mu}\,.
\end{align}
For $\left\langle \br^2 z\text{u}_z\right\rangle_{s}$,
\begin{align}
&\left[s+2 D_r+3\mu\right]\left\langle \br^2z\text{u}_{z}\right\rangle_{s}=  10  D\left\langle z \text{u}_{z}\right\rangle_{s}+2 v_0\left\langle\br \cdot \bu z\text{u}_{z}\right\rangle_{s}+v_0\left\langle \br^2 \text{u}_{z}^{2}\right\rangle_{s}\,.
\end{align}
For $\left\langle\br \cdot \bu z\text{u}_{z}\right\rangle_{s}$,
\begin{align}
&\left[s+6D_r +2\mu+\frac{\omega^{2}}{s+6 D_r+2\mu}\right]\left\langle\br \cdot \bu z \text{u}_{z}\right\rangle_{s}=2 D\left\langle \text{u}_{z}^{2}\right\rangle_{s}+2 D_r\left\langle z^{2}\right\rangle_{s}+v_0\left\langle z \text{u}_{z}\right\rangle_{s}\nonumber\\
&+\frac{\omega^{2}\left\langle z^{2} \text{u}_{z}^{2}\right\rangle_{s}}{s+6 D_r+2\mu}+v_0\left[\left\langle\br \cdot \bu \text{u}_{z}^{2}\right\rangle_{s}+\frac{\omega\left\langle R_{z}\left(\br \cdot \bu \text{u}_{z}^{2})\right\rangle_{s}\right.}{s+6 D_r+2\mu}\right]\,,
\end{align}
and
\begin{align}
&\left(s+6 D_r+2\mu\right)\left\langle R_{z}\left(\br\cdot \bu z \text{u}_{z}\right)\right\rangle_{s}=v_0\left\langle R_{z}\left(\br\cdot \bu \text{u}_{z}^{2}\right)\right\rangle_{s}  -\omega\left\langle\br\cdot \bu z \text{u}_{z}\right\rangle_{s}+\omega\left\langle z^{2} \text{u}_{z}^{2}\right\rangle_{s}\,.
\end{align}
For $\left\langle\br \cdot \bu \text{u}_{z}^{2}\right\rangle_{s}$,
\begin{align}
&\left[s+12 D_r+\mu+\frac{\omega^{2}}{s+12 D_r+\mu}\right] \left\langle\br\cdot \bu \text{u}_{z}^{2}\right\rangle_{s}=  2 D_r\langle\br\cdot \bu\rangle_{s}+4 D_r\left\langle z \text{u}_{z}\right\rangle_{s}+v_0\left\langle \text{u}_{z}^{2}\right\rangle_{s} \nonumber\\
&+\frac{\omega^{2}\left\langle z \text{u}_{z}^{3}\right\rangle_{s}}{s+12D_r+\mu}+\frac{2 \omega D_r\left\langle R_{z}(\br\cdot \bu)\right\rangle_{s} }{s+12 D_r+\mu}\,,
\end{align}
and
\begin{align}
{\left[s+12 D_r+\mu\right]\left\langle R_{z}\left(\br\cdot \bu \text{u}_{z}^{2}\right)\right\rangle_{s} = 2 D_r\left\langle R_{z}(\br\cdot \bu)\right\rangle_{s} } &-\omega\left\langle(\br\cdot \bu) \text{u}_{z}^{2}\right\rangle_{S}
+\Omega\left\langle  z \text{u}_{z}^{3}\right\rangle_{s}\,.
\end{align}
Now,
\begin{align}
\left[s+12D_r+\mu\right]\left\langle z \text{u}_{z}^{3}\right\rangle_{s}=6 D_r\left\langle z \text{u}_{z}\right\rangle_{s}+v_0\left\langle \text{u}_{z}^{4}\right\rangle_{s}\,,
\end{align}
with
$$
\nonumber
\left[s+20 D_r\right]\left\langle \text{u}_{z}^{4}\right\rangle_{s}=12 D_r\left\langle \text{u}_{z}^{2}\right\rangle_{s}+\text{u}_{z0}^4\,,
$$
\begin{align}
\left(s+6 D_r+2 \mu\right)\left\langle z^{2} \text{u}_{z}^{2}\right\rangle_{s}=2 D\left\langle \text{u}_{z}^{2}\right\rangle_{s}+2 D_r \left\langle z^{2}\right\rangle_{s}+2 v_0\left\langle z \text{u}_{z}^{3}\right\rangle_{s}\,,
\end{align}
\begin{align}
\left(s+6 D_r+2 \mu\right)\left\langle \br^2 \text{u}_{z}^{2}\right\rangle_{s}=2 D_r\left\langle \br^2\right\rangle_{s}+6 D \left\langle \text{u}_{z}^{2}\right\rangle_{s}+2 v_0\left\langle\br \cdot \bu \text{u}_{z}^{2}\right\rangle_{s}\,.
\end{align}

\noindent
{\bf{Calculation of $\left\langle(\br \cdot \bu)^{2}\right\rangle_{s}$}}

\begin{align}
&\left[s+6 D_r+ 2\mu+\frac{4 \omega^{2}}{s+6 D_r +2\mu}\right]\left\langle(\br\cdot \bu)^{2}\right\rangle_{s}= \frac{2 D}{s}+2 D_r\left\langle \br^2\right\rangle_{s}+2 v_0\langle\br\cdot \bu\rangle_{s} \nonumber\\
&+\frac{2 \omega v_0\left\langle R_{z}(\br\cdot \bu)\right\rangle_{s}}{s+6 D_r+2\mu} +6 \omega^{2} \frac{\left\langle\br \cdot \bu z \text{u}_{z}\right\rangle_{s}}{s+6 D_r +2\mu}+2 \omega^{2} \frac{\left\langle \br^2\right\rangle_{s}}{s+6 D_r+2\mu}- \frac{2 \omega^{2}\left\langle z^{2}\right\rangle_{s}}{s+6 D_r +2\mu} \nonumber\\
&-\frac{2 \omega^{2}\left\langle \br^2 \text{u}_{z}^{2}\right\rangle_{s}}{s+6 D_r +2\mu}\,,
\end{align}
again,
\begin{align}
&{\left(s+6 D_r +2\mu\right) \left\langle R_{z}(\br\cdot \vec{u})^{2}\right\rangle_{s} } =2 v_0\left\langle R_{z}(\br\cdot \bu)\right\rangle_{s}-4 \omega\left\langle(\br\cdot \bu)^{2}\right\rangle_{s} +6 \omega\left\langle\br\cdot \bu z \text{u}_{z}\right\rangle _{s}\nonumber\\ 
&+2 \omega\left\langle \br^2\right\rangle_{s}-2 \omega\left\langle z^{2}\right\rangle_{s} -2 \omega\left\langle \br^2 \text{u}_{z}^{2}\right\rangle_{s}\,. 
\end{align}
Using all the expressions, we can calculate the steady-state expression of the fourth moment of the displacement by means of the final value theorem. The dimensionless expression of the fourth moment of the displacement,
\begin{align}
&\langle\tbr^4\rangle_{\rm st}= \frac{15}{\beta ^2}+\frac{10 \text{Pe}^2 \left(3 (\beta +2)^2+\Omega ^2\right)}{3 \beta ^2 (\beta +2) \left((\beta +2)^2+\Omega ^2\right)} 
\nonumber\\&+\text{Pe}^4 \Bigg(\frac{4 (\beta +2) (\beta +3)^3 (3 \beta +2) (3 \beta +5)}{\beta ^2 \left((\beta +2)^2+\Omega ^2\right) \left((\beta +3)^2+\Omega ^2\right) \left(4 (\beta +3)^2+\Omega ^2\right) \left((3 \beta +2)^2+\Omega ^2\right)}
\nonumber\\&+\frac{(\beta +3) (\beta  (\beta  (\beta  (3 \beta  (111 \beta +905)+8632)+13104)+9424)+2640) \Omega ^2}{3 \beta ^2 (\beta +2) (3 \beta +2) \left((\beta +2)^2+\Omega ^2\right) \left((\beta +3)^2+\Omega ^2\right) \left(4 (\beta +3)^2+\Omega ^2\right) \left((3 \beta +2)^2+\Omega ^2\right)}
\nonumber\\&
+\frac{(\beta  (\beta  (\beta  (\beta  (759 \beta +7445)+27802)+49370)+42644)+15060) \Omega ^4}{15 \beta ^2 (\beta +2) (\beta +3) (3 \beta +2) \left((\beta +2)^2+\Omega ^2\right) \left((\beta +3)^2+\Omega ^2\right) \left(4 (\beta +3)^2+\Omega ^2\right) \left((3 \beta +2)^2+\Omega ^2\right)}
\nonumber\\&
+\frac{3 (3 \beta +5) \Omega ^8+(\beta  (\beta  (147 \beta +755)+1367)+875) \Omega ^6}{15 \beta ^2 (\beta +2) (\beta +3) (3 \beta +2) \left((\beta +2)^2+\Omega ^2\right) \left((\beta +3)^2+\Omega ^2\right) \left(4 (\beta +3)^2+\Omega ^2\right) \left((3 \beta +2)^2+\Omega ^2\right)}\Bigg)\,.
\label{eq:r4_3d_st_dimensionless}
\end{align}

\subsection{\bf Excess kurtosis}
We can calculate excess kurtosis by using Eq.~\eqref{ex_kurt}.
Averaging over the initial orientations, Eq.~\eqref{mu4_avg} gives,
\begin{align}
&\tilde \mu_4  = \langle\tbr^2\rangle^2+2\Big( \langle  \tilde x^2 \rangle^2 +\langle  \tilde y^2 \rangle^2+ \langle  \tilde z^2 \rangle^2 +\langle  \tilde x \tilde y \rangle^2+\langle  \tilde y \tilde z \rangle^2+\langle  \tilde z \tilde x \rangle^2\Big)\,.
\label{eq:kurtosis_3d_first}
\end{align}
The steady-state expressions of the moments in the long time limit are obtained by applying the final value theorem to their Laplace transformed forms;
\begin{align} 
 \langle \tilde{x}^2\rangle_{\rm st}& = \frac{\Pe^2 (2+\beta )}{3 \beta  \left((2 +\beta )^2+\Omega ^2\right)}+\frac{1}{\beta }\,,
 \\ \langle \tilde{y}^2\rangle_{\rm st} &= \frac{\Pe^2 (2+\beta )}{3 \beta  \left((2 +\beta )^2+\Omega ^2\right)}+\frac{1}{\beta }\,,
 \\\langle \tilde{z}^2\rangle_{\rm st} &= \frac{1}{\beta}+\frac{\Pe^2}{3\beta(2 +\beta)}\,, 
 \\ \langle \tilde x \tilde y\rangle_{\rm st}&=\langle \tilde x \tilde z\rangle_{\rm st}= \langle \tilde y \tilde z\rangle_{\rm st}=0\,.
\end{align}
The dimensionless form of the steady-state excess kurtosis,

\begin{align}
    \tilde{\mathcal{K}}_{\rm st} = \frac{\mathcal{A}}{\mathcal{B}} -1\,.
\end{align}

\noindent 
Where:

\begin{align*}
    \mathcal{A} &= 3 \left[(\beta + 2)\Omega^2 + (\beta + 2)^3\right]^2 \times
    \Bigg[
    225 + \frac{\text{Pe}^2}{(\beta + 2) \left[(\beta + 2)^2 + \Omega^2\right]} \Bigg( 
    50\left[3(\beta + 2)^2 + \Omega^2\right] \\
    &\quad + \text{Pe}^2 \Big\{60(\beta + 2)^2(\beta + 3)^4(3\beta + 2)^2(3\beta + 5) \\
    &\quad + 3(3\beta + 5)\Omega^8 + (\beta (\beta (147\beta + 755) + 1367) + 875)\Omega^6 \\
    &\quad + (\beta(\beta(\beta(\beta(759\beta + 7445) + 27802) + 49370) + 42644) + 15060)\Omega^4 \\
    &\quad + 5(\beta + 3)^2(\beta(\beta(\beta(3\beta(111\beta + 905) + 8632) + 13104) + 9424) + 2640)\Omega^2
    \Big\}
    \Bigg)
    \Bigg]\,.
\end{align*}

\begin{align*}
    \mathcal{B} &= 50(\beta + 2)^2 \Omega^2 \left(3\beta + \text{Pe}^2 + 6\right) \left(9(\beta + 2) + \text{Pe}^2\right) + 75(\beta + 2)^4 \left(3\beta + \text{Pe}^2 + 6\right)^2 \\
    &\quad + 15\Omega^4 \left[45(\beta + 2)^2 + \text{Pe}^4 + 10(\beta + 2)\text{Pe}^2\right]\,.
\end{align*}
\noindent
\textbf{Limiting cases:} For high and low values of the parameters, the steady-state excess kurtosis evolves as,
\noindent
{\allowdisplaybreaks
\begin{align}
    \begin{aligned}
        \lim_{\text{Pe}\to 0} \tilde {\cal{K}}_{\rm st} =& -\Pe^4 \frac{\mathcal{A}_1}{\mathcal{B}_1}+O(\Pe^6)\,, 
        \\
        \lim_{\text{Pe}\to \infty} \tilde {\cal{K}}_{\rm st} =& -1+\frac{\mathcal{A}_2}{\mathcal{B}_2}+O\left(\frac{1}{\Pe^2}\right)\,,
    \\
         \lim_{\beta\to 0} \tilde {\cal{K}}_{\rm st} =& -\beta\frac{\mathcal{A}_3}{\mathcal{B}_3}
         +O\left(\beta ^2\right)\,,
         \\
        \lim_{\beta\to \infty} \tilde {\cal{K}}_{\rm st} =& -\frac{2 \text{Pe}^4}{45 \beta ^2}+\frac{4 \text{Pe}^4 \left(\text{Pe}^2+6\right)}{135 \beta ^3}-\frac{2 \left(\text{Pe}^4 \left(\text{Pe}^4+12 \text{Pe}^2-4 \Omega ^2+30\right)\right)}{135 \beta ^4}+O\left(\frac{1}{\beta ^5}\right)\,,
    \\
        \lim_{\Omega\to 0} \tilde {\cal{K}}_{\rm st} =&-\frac{2 \left(\beta  (3 \beta +11) \text{Pe}^4\right)}{5 \left((\beta +3) (3 \beta +2) \left(3 \beta +\text{Pe}^2+6\right)^2\right)}+O\left(\Omega ^2\right)\,,
        \\
        \lim_{\Omega\to \infty} \tilde {\cal{K}}_{\rm st} =& -\frac{2 \left(\beta  (3 \beta +11) \text{Pe}^4\right)}{5 \left((\beta +3) (3 \beta +2) \left(45 (\beta +2)^2+\text{Pe}^4+10 (\beta +2) \text{Pe}^2\right)\right)}+O\left(\frac{1}{\Omega ^2}\right)\,.
    \end{aligned}
\end{align}}
where
\allowdisplaybreaks{\begin{align*}
&\mathcal{A}_1=2\beta\Bigg[60 (\beta +2)^4 (\beta +3)^4 (3 \beta +2)^2 (3 \beta +11)
\\&+5 (\beta +2)^2 (\beta +3)^2 (\beta  (\beta  (3 \beta  (\beta  (219 \beta +1955)+6432)+28444)+17952)+4320) \Omega ^2
\\&+\bigg(2424 \beta ^7+35054 \beta ^6+211985 \beta ^5+691419 \beta ^4+1307668 \beta ^3+1433620 \beta ^2+856368 \beta +228192\bigg) \Omega ^4
\\&+\left(\beta  \left(\beta  \left(\beta  \left(906 \beta ^2+9160 \beta +36133\right)+70107\right)+69008\right)+30344\right) \Omega ^6
\\&+(\beta  (2 \beta  (78 \beta +499)+2135)+1841) \Omega ^8+3 (3 \beta +11) \Omega ^{10}\Bigg]\,,
\\&\mathcal{B}_1=675 (\beta +3) (3 \beta +2) \left((\beta +3)^2+\Omega ^2\right) \left(4 (\beta +3)^2+\Omega ^2\right) \left((3 \beta +2)^2+\Omega ^2\right) \left((\beta +2) \Omega ^2+(\beta +2)^3\right)^2\,,\\ \\
&\mathcal{A}_2= 3 \left((\beta +2) \Omega ^2+(\beta +2)^3\right)^2\Bigg[60 (\beta +2)^2 (\beta +3)^4 (3 \beta +2)^2 (3 \beta +5)
\\&+5 (\beta +3)^2 \Big(\beta  (\beta  (\beta  (3 \beta  (111 \beta +905)+8632)+13104)+9424)+2640\Big) \Omega ^2
\\&+(\beta  (\beta  (\beta  (\beta  (759 \beta +7445)+27802)+49370)+42644)+15060) \Omega ^4
\\&+(\beta  (\beta  (147 \beta +755)+1367)+875) \Omega ^6+3 (3 \beta +5) \Omega ^8\Bigg]\,,
\\
&\mathcal{B}_2=5 (\beta +2) (\beta +3) (3 \beta +2) \left((\beta +2)^2+\Omega ^2\right) \left((\beta +3)^2+\Omega ^2\right) \left(4 (\beta +3)^2+\Omega ^2\right)
\\&\times\left((3 \beta +2)^2+\Omega ^2\right) \left(10 (\beta +2)^2 \Omega ^2+15 (\beta +2)^4+3 \Omega ^4\right)\,,\\ \\
&\mathcal{A}_3=\text{Pe}^4 \left(33 \Omega ^{10}+1841 \Omega ^8+30344 \Omega ^6+228192 \Omega ^4+777600 \Omega ^2+3421440\right)\,,\\
&\mathcal{B}_3= 15 \left(\Omega ^2+4\right) \left(\Omega ^2+9\right) \left(\Omega ^2+36\right) 
\\&\times\left(40 \left(\text{Pe}^2+6\right) \left(\text{Pe}^2+18\right) \Omega ^2+240 \left(\text{Pe}^2+6\right)^2+3 \left(\text{Pe}^4+20 \text{Pe}^2+180\right) \Omega ^4\right)\,.
\end{align*}}
\end{widetext}

\subsection*{References}
\bibliography{reference} 

\end{document}